\documentclass[twocolumn,showpacs,preprintnumbers,superscriptaddress,%
 amsmath,amssymb,floatfix, pra]{revtex4}

\usepackage{graphicx}
\usepackage{bm}
\usepackage{bbm}
\usepackage{subfigure}
\usepackage{color}
\usepackage{timenow}
\newcommand{\mri}{\mathrm{i}}
\newcommand{\Exp}[1]{\mathrm{e}^{\mbox{\footnotesize$#1$}}}
\renewcommand{\vec}[1]{\bm{#1}}

\begin{document}
\title{Ultracold Fermions in a Graphene-Type Optical Lattice}



\author{Kean Loon \surname{Lee} }
\email{leekeanl@sps.nus.edu.sg}
\affiliation{Centre for Quantum Technologies, %
National University of Singapore, 3 Science Drive 2, Singapore 117543, %
Singapore}
\affiliation{Laboratoire Kastler Brossel, Ecole Normale Sup\'erieure, CNRS, %
UPMC; 4 Place Jussieu, 75005 Paris, France}
\affiliation{NUS Graduate School for Integrative Sciences and Engineering, %
National University of Singapore, Singapore}

\author{Beno\^it \surname{Gr\'emaud}}
\affiliation{Laboratoire Kastler Brossel, Ecole Normale Sup\'erieure, CNRS, %
UPMC; 4 Place Jussieu, 75005 Paris, France}
\affiliation{Centre for Quantum Technologies, %
National University of Singapore, 3 Science Drive 2, Singapore 117543, %
Singapore}
\affiliation{Department of Physics, National University of Singapore, %
2 Science Drive 3, Singapore 117542, Singapore}

\author{Rui \surname{Han}}
\affiliation{Centre for Quantum Technologies, %
National University of Singapore, 3 Science Drive 2, Singapore 117543, %
Singapore}

\author{Berthold-Georg \surname{Englert}}
\affiliation{Centre for Quantum Technologies, %
National University of Singapore, 3 Science Drive 2, Singapore 117543, %
Singapore}
\affiliation{Department of Physics, National University of Singapore, %
2 Science Drive 3, Singapore 117542, Singapore}

\author{Christian \surname{Miniatura}}
\affiliation{Institut Non Lin\'eaire de Nice, UMR 6618, UNS, CNRS; %
1361 route des Lucioles, 06560 Valbonne, France}
\affiliation{Centre for Quantum Technologies, %
National University of Singapore, 3 Science Drive 2, Singapore 117543, %
Singapore}
\affiliation{Department of Physics, National University of Singapore, %
2 Science Drive 3, Singapore 117542, Singapore}

\date{\textbf{\Now}}

\begin{abstract}
Some important features of the graphene physics can be reproduced by loading ultracold fermionic atoms in a two-dimensional optical lattice with honeycomb symmetry and we address here its experimental feasibility. We analyze in great details the optical lattice generated by the coherent superposition of three coplanar running laser waves with respective angles $2\pi/3$. The corresponding band structure displays Dirac cones located at the corners of the Brillouin zone and close to half-filling this system is well described by massless Dirac fermions. We characterize their properties by accurately deriving the nearest-neighbor hopping parameter $t_0$ as a function of the optical lattice parameters. Our semi-classical instanton method proves in excellent agreement with an exact numerical diagonalization of the full Hamilton operator in the tight-binding regime. We conclude that the temperature range needed to access the Dirac fermions regime is within experimental reach. We also analyze imperfections in the laser configuration as they lead to optical lattice distortions which affect the Dirac fermions. We show that the Dirac cones do survive up to some critical intensity or angle mismatches which are easily controlled in actual experiments. In the tight-binding regime, we predict, and numerically confirm, that these critical mismatches are inversely proportional to the square-root of the optical potential strength. We also briefly discuss the interesting possibility of fine-tuning the mass of the Dirac fermions by controlling the laser phase in an optical lattice generated by the incoherent superposition of three coplanar independent standing waves with respective angles $2\pi/3$.
\end{abstract}

\begin{widetext}
\maketitle
\end{widetext}

\section{Introduction}\label{sec:intro}

In 2004, researchers in Manchester isolated one-atom thick sheets of carbon atoms, with the atoms organized in a planar honeycomb structure \cite{novoselov04}. Such a material is referred to as graphene and is of utmost importance in condensed-matter physics since by stacking it one gets the graphite structure, and by wrapping it one gets carbon nanotubes and fullerenes \cite{saito98}. Graphene is also of great theoretical interest because it provides a physical realization of two-dimensional field theories with quantum anomalies \cite{semenoff84}. Indeed, the effective theory that describes the low-energy electronic excitations in graphene is that of two-dimensional massless Weyl-Dirac fermions. In graphene these massless fermions propagate with about one 300th of the speed of light. Triggered by the Manchester discovery, an intense activity has flourished in the field, and continues to flourish, as witnessed by Refs.~\cite{wilson06,neto06,zhou06,katsnelson06,katsnelson07,neto09}, for example. The reported and predicted phenomena include the Klein paradox (the perfect transmission of relativistic particles through high and wide potential barriers) \cite{katsnelson06}, the anomalous quantum Hall effect induced by Berry phases \cite{novoselov2005, zhang2005}, and its corresponding modified Landau levels \cite{li2007}.

It is now well established that some condensed-matter phenomena can be
reproduced by loading ultracold atoms into optical lattices
\cite{bloch05,lewenstein07}.
The great advantage is that the relevant parameters are accessible for
accurate
control (shape and strength of the light potential, atom-atom interaction
strength via Feshbach resonances \cite{timmermans99}, etc.) while spurious
effects that destroy the quantum coherence are absent, such as the analog of
the electron-phonon interaction.
Our present objective is to analyze a scheme capable of reproducing in atomic
physics the unique situation found in graphene \cite{zhu07}.
It consists of creating a two-dimensional honeycomb optical lattice and
loading it with ultracold fermions like the neutral Lithium-6 or Potassium-40
atoms.

Parts of this paper recall known results.
In addition to the need of setting the stage and introducing the notational
conventions, there is also the intention to bridge the solid-state community
and the atomic physics community on the particular subject of massless Dirac
fermions as observed in graphene sheets and its counterpart in atomic physics.
We also present extensions of previous solid-state works in the atomic
physics context and report a number of new results.

We analyze the various experimental parameters that need to be controlled
in order to reproduce, with cold atoms trapped in an optical lattice,
the physics at work in graphene.
After briefly introducing optical lattices, we first explain how to create an
optical lattice with the honeycomb symmetry and analyze its crystallographic
features.
We then calculate the band structure in the tight-binding approximation and by
exact diagonalization, thereby providing evidence for the occurrence of the
so-called Dirac points.
Next, we evaluate the nearest-neighbors hopping amplitude by using a semi-classical instanton
method.
For the benefit of possible experiments we give the necessary requirements for
reaching the massless Dirac fermions regime.
Finally, we examine how massless Dirac fermions survive lattice distortions
that could result from intensity-unbalanced or misaligned laser beams.
These distortions open the way to new physics related to the quantum Hall
effect \cite{dietl08}.
We will close by briefly mentioning possible experiments to target for
noninteracting and interacting ultracold fermions \cite{paiva05, zhao06}.

\section{The honeycomb optical lattice}\label{sec:perfect}
\subsection{Radiative forces and optical lattices}\label{sec:forces}

A two-level atom (with angular frequency separation
$\omega_{at}$ and excited-state angular frequency width $\Gamma$)
that interacts with a monochromatic laser field with complex amplitude
$\vec{\mathcal{E}}(\vec{r},t) = \vec{E}(\vec{r})
\,\Exp{-\mri\omega_L t}$ gets polarized and experiences radiative forces due
to photon absorption and emission cycles \cite{RedBookCCT, wallis95}.
When the light frequency is tuned far away from the atomic resonance,
i.e., when the light detuning $\delta =\omega_L -\omega_{at}$ is much larger than
$\Gamma$, the field-induced saturation effects are negligible and the atom
essentially keeps staying in its ground state.
In this situation, the atom-field interaction is dominated by stimulated
emission processes where the atomic dipole absorbs a photon from one Fourier
component of the field and radiates it back into the same or another one of
these Fourier modes.
In each such stimulated cycle, there is a momentum transfer to the atom and,
as a net result, the atom experiences an average force in the course of time.
This dipole force exerted by the field onto the atom in its ground state is
conservative.
It derives from the polarization energy shift of the atomic levels (AC Stark
or light shifts) \cite{grimm00} and the dipole potential $V(\bm{r})$ is given
by
\begin{equation}
\label{DipPot}
V(\vec{r}) = \frac{\hbar \Gamma}{8} \,\frac{\Gamma}{\delta}
            \,\frac{I(\vec{r})}{I_s}\,,
\end{equation}
where $I(\vec{r}) = \epsilon_0 c \biglb|\vec{E}(\vec{r})\bigrb|^2 /2$
is the light field intensity (time-averaged energy current density) at the
center-of-mass position $\bm{r}$ of the atom and $I_s$ is the saturation intensity
of the atom under consideration.

For multi-level atoms, the situation is more complicated as the dipole
potential now depends on the particular atomic ground state sub-level under
consideration.
However, if the laser detuning $\delta$ is much larger than the fine and
hyperfine structure splittings of the atomic electronic transition, then all
ground state atomic sub-levels will essentially experience the same dipole
potential.
This common potential turns out to be given by \eqref{DipPot} as well.
Hence, by conveniently tailoring the space and time dependence of the laser
field, one can produce a great variety of dipole potentials and thus
manipulate the ground state atomic motion.

Optical lattices are periodic intensity patterns of light obtained through the
interference of several monochromatic laser beams \cite{grynberg2001}.
By loading ultracold atoms into such artificial crystals of light one obtains
periodic arrays of atoms.
Indeed, as seen from \eqref{DipPot}, when the light field is blue-detuned from
the atomic resonance ($\delta >0$), then the atoms can be trapped in the
field-intensity minima whereas for red-tuned light ($\delta <0$) they can be
trapped at the field intensity maxima.
Such arrays of ultracold atoms trapped in optical lattices have been used in a
wide variety of experiments.
As recently evidenced by the observation of the Mott-Hubbard transition with
degenerate gases \cite{greiner02}, they have proven to be a unique tool to
mimic, test and go beyond phenomena observed until now in the condensed-matter
realm \cite{jaksch98, lewenstein07}.
They also have a promising potential for the implementation of quantum
simulators and for quantum information processing purposes
\cite{jaksch99,bloch05, chen09}.

\subsection{Optical lattice with honeycomb structure}
\label{sec:optlattice}

\subsubsection{Field configuration and associated dipole potential}

\begin{figure}[tb]
\includegraphics[scale=0.7]{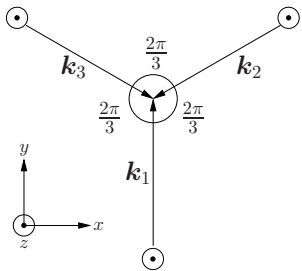}
\caption{\label{fig:threeBeam} The coplanar three-beam configuration used to
  generate the honeycomb lattice. All beams have the same frequency, strength
  and linear polarization orthogonal to their common propagation plane. The
  honeycomb lattice under consideration is obtained for blue-detuned beams
  with respective angles $2\pi/3$.
  For these symmetric laser beams, the time-averaged radiation pressure ---
  albeit small at large detuning --- vanishes in this configuration.
  By reversing the propagation direction of one of the lasers, such that
  ${\vec{k}_1=\vec{k}_2+\vec{k}_3}$, say, a triangular lattice of a different
  geometry is formed.
  We will, however, exclusively deal with the
  ${\vec{k}_1+\vec{k}_2+\vec{k}_3=0}$ case.
  }
\end{figure}

The simplest possible optical lattice with honeycomb structure is generated by
superposing three coplanar traveling plane waves that have
the same angular frequency ${\omega_L=ck_L}$, the same field strength $E_0>0$, the same polarization and the three wave vectors $\vec{k}_a$ form a
trine: their sum vanishes and the angle between any two of them is $2\pi/3$,
\begin{equation}
  \label{eq:ktrine}
  \vec{k}_1+\vec{k}_2+\vec{k}_3=0\,,\quad
  \vec{k}_a\cdot\vec{k}_b
   =k_L^2\Bigl(\frac{3}{2}\delta_{ab}-\frac{1}{2}\Bigr)
\end{equation}
with $a,b=1,2,3$ and $\delta_{ab}$ is the Kronecker symbol \cite{grynberg2001}.
As is illustrated in Fig.~\ref{fig:threeBeam}, we choose the $x,y$-plane as
the common plane of propagation and, to be specific, use
\begin{equation}
  \label{eq:kvectorsym}
\vec{k}_1=k_L\vec{e}_y\,,\quad\left.
  \begin{array}{c}
    \vec{k}_2\\[0.5ex]\vec{k}_3
  \end{array}\right\}=k_L\frac{\mp\sqrt{3}\vec{e}_x-\vec{e}_y}{2}
\end{equation}
for the parameterization of the wave vectors.

Further, we take all fields to be linearly polarized orthogonal to the plane,
so that the three complex field amplitudes are given by
\begin{equation}\label{eq:trinefields}
	\vec{\mathcal{E}}_a(\vec{r},t) = E_0\,
        \Exp{\mri(\vec{k}_a\cdot\vec{r}-\phi_a)}
        \Exp{-\mri\omega_L t}\,\vec{e}_z
\end{equation}
where $\phi_a$ is the phase of the $a$th field for $t=0$ at
$\vec{r}=0$.
We note that a joint shift of the reference points in time and space,
\begin{equation}
  \label{eq:jointshift}
  t\to t-\frac{1}{3\omega_L}\sum_a\phi_a\,,\quad
  \vec{r}\to\vec{r}+\frac{2}{3k_L^2}\sum_a\phi_a\vec{k}_a\,,
\end{equation}
removes the phases $\phi_a$ from (\ref{eq:trinefields}), so that the simple
choice ${\phi_1=\phi_2=\phi_3=0}$ is permissible, and we adopt this
convention.
In an experimental implementation, one would need to stabilize the phase
differences $\phi_a-\phi_b$  to prevent a rapid jitter of the lattice that
could perturb the atoms trapped in the potential minima.

The dipole potential \eqref{DipPot} generated by the electric field
$\vec{E}=\sum_a\vec{\mathcal{E}}_a$ is of the form
\begin{equation}\label{eq:HCpotential}
  V(\vec{r})=V_0\biglb|f(\vec{r})\bigrb|^2=V_0v(\vec{r})\quad\mbox{with}\enskip
       V_0=\frac{\hbar \Gamma}{8}\frac{\Gamma}{\delta}\frac{I_0}{I_s}\,,
\end{equation}
where $I_0$ is the intensity associated with the field strength $E_0$. The total dimensionless field amplitude $f(\vec{r})$ and the dimensionless optical potential $v(\vec{r})$ are given by
\begin{equation}\label{eq:fieldamp}
f(\vec{r}) = 1 + \exp(-\mri \vec{b}_1\cdot\vec{r})+ \exp(\mri \vec{b}_2\cdot\vec{r})
\end{equation}
and
\begin{equation}
\label{eq:honeypot}
v(\vec{r})=3+2\cos(\vec{b}_1\cdot\vec{r})+2\cos(\vec{b}_2\cdot\vec{r})+2\cos\left((\vec{b}_1+\vec{b}_2)\cdot\vec{r}\right)\,,
\end{equation}
where ${\vec{b}_1 =\vec{k}_3-\vec{k}_1}$ and ${\vec{b}_2 =\vec{k}_1-\vec{k}_2}$ feature the reciprocal primitive vectors. For the parameterization \eqref{eq:kvectorsym}, we have
\begin{equation}
  \label{eq:bvectors}
  \left.\begin{array}{c}
     \vec{b}_1\\[0.5ex]\vec{b}_2
  \end{array}\right\}=
  \kappa\frac{\vec{e}_x\mp\sqrt{3}\vec{e}_y}{2}
\end{equation}
with $\kappa=\biglb|\vec{b}_a\bigrb|=\sqrt{3}k_L$. One may further notice that the periodic patterns associated to each of the cosine terms in
\eqref{eq:honeypot} have the same spatial period of $(2\pi/k_L)/\sqrt{3}$, about 58\%
of the laser wavelength $\lambda_L=2\pi/k_L$.

Linear combinations of the Brillouin vectors with integer coefficients define
the reciprocal lattice $\tilde{\mathcal{B}}$, a regular pattern in $\vec{k}$-space,
\begin{equation}
  \label{eq:recilat1}
\tilde{\mathcal{B}}=
\bigl\{n_1\vec{b}_1+ n_2\vec{b}_2\bigm| n_1,n_2=0,\pm1,\pm2,\dots \bigr\}\,.
\end{equation}
The reciprocal lattice is central to all studies of the dynamics of particles
that move under the influence of the given periodic potential~\cite{ashcroft76}.

\begin{figure}[tb]
\centerline{\includegraphics[scale=1.0]{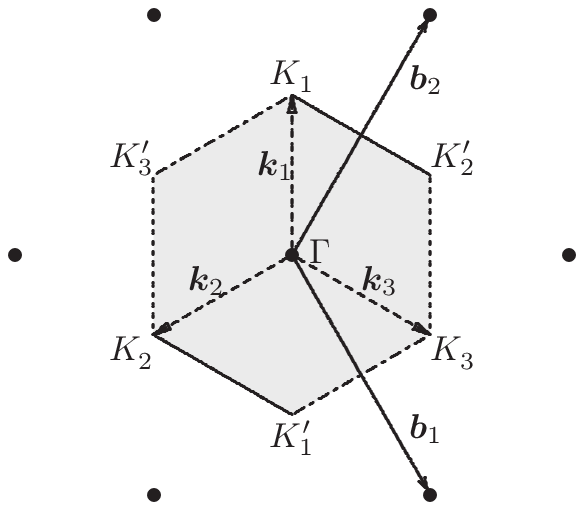}\rule[-10pt]{0pt}{20pt}}
\caption{\label{fig:honeycombRecipLattice}
The triangular reciprocal lattice $\tilde{\mathcal{B}}$ associated with the
triangular Bravais lattice of Fig.~\ref{fig:ABChexagon}.
It is spanned by the reciprocal primitive vectors $\vec{b}_1$ and $\vec{b}_2$ of \eqref{eq:bvectors},
and is also a triangular lattice (as indicated by the full dots).
The shaded region identifies the first Brillouin zone $\Omega$ which is here a
regular hexagon.
Its center is conventionally named $\Gamma$ in the solid-state literature.
Opposite edges are in fact identical as they only differ by a translation in
the reciprocal lattice.
This feature is emphasized by drawing the identical edges with the same
(solid, dashed or dash-dotted) line.
For the same reason, the three corners $K_a$  ($a=1,2,3$) are to be identified
with each other, and likewise the three corners $K'_a$ are really only one
point in $\Omega$.
Thus only two of the six corners, collectively labeled as $K$ and $K'$ and
known as the Dirac points, are different. Also shown are the wave vectors of the three coplanar plane waves (dashed arrows).
}
\end{figure}

In particular, one domain in reciprocal space of utmost importance is the first Brillouin zone $\Omega$, defined as the so-called primitive Wigner-Seitz cell \cite{ashcroft76} of $\tilde{\mathcal{B}}$, see Fig.~\ref{fig:honeycombRecipLattice}. It is a \emph{regular} hexagon but with the subtle feature that opposite edges
are to be identified with each other since they can be related by a displacement vector in
$\tilde{\mathcal{B}}$. By the same token the three corners $K_a$ (respectively $K'_a$) have to be
identified with one another and we collectively denote them by $K$ (respectively $K'$).
These two different corners $K$ and $K'$ are known in the graphene literature
as the Dirac points for a reason that will become clear in the next section.
Upon denoting $\vec{K} \equiv \vec{K}_1$ and $\vec{K}' \equiv \vec{K}'_1$,
their positions in $\Omega$ are given by the wave vector of the lasers that
generate the optical honeycomb potential,
\begin{equation}
\label{Kvectors}
\vec{K} = - \vec{K}'= \frac{1}{3}(\vec{b}_2-\vec{b}_1)=\vec{k}_1
\end{equation}
and ${\vec{K}_2=\vec{k}_2 = \vec{K}-\vec{b}_2}$,
${\vec{K}_3=\vec{k}_3 = \vec{K}+\vec{b}_1}$, as well as
$\vec{K_a}=-\vec{K}'_a$.

\subsubsection{Triangular Bravais lattice}

The dimensionless potential \eqref{eq:honeypot} consists of a periodic
two-dimensional array of maxima, minima, and saddle points,
generated by repeated translations of a primitive unit tile called the
\emph{basis}.
The underlying lattice geometry itself is encapsulated in the associated
Bravais lattice $\mathcal{B}$, that is
\begin{eqnarray}
  \label{eq:bravilat}
\mathcal{B}&=&
\bigl\{m_1 \vec{a}_1 + m_2\bm{a}_2 \bigm| m_1,m_2=0,\pm1,\pm2,\dots \bigr\},
\end{eqnarray}
such that the value of the potential is not affected by any displacement $\vec{R}\in \mathcal{B}$, $v(\vec{r}+\vec{R})=v(\vec{r})$.

The Bravais primitive vectors $\vec{a}_a$ are constructed based on the relation
\begin{equation}
  \label{eq:orthrel}
  \vec{a}_a\cdot\vec{b}_b = 2\pi\delta_{ab}.
\end{equation}
In other words, the Bravais lattice $\mathcal{B}$ and the Brillouin lattice
$\tilde{\mathcal{B}}$ constitute dual spaces. Supplementing \eqref{eq:bvectors}, we have the
explicit parameterization
\begin{equation}
  \label{eq:avectors}
  \left.\begin{array}{c}
    \vec{a}_1\\[0.5ex]\vec{a}_2
  \end{array}\right\}=\Lambda\frac{\sqrt{3}\vec{e}_x\mp\vec{e}_y}{2}
\end{equation}
where ${\Lambda=\biglb|\vec{a}_a\bigrb|=4\pi/(3k_L)=2\lambda_L/3}$ is the common length of the Bravais primitive vectors.

\begin{figure}[tb]
	\includegraphics[scale=1.25]{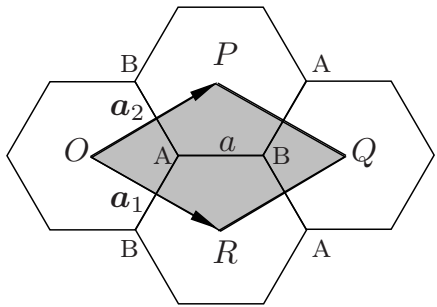}\caption{\label{fig:basis} The underlying Bravais lattice $\mathcal{B}$ of a two-dimensional honeycomb is the two-dimensional triangular Bravais lattice with a two-point basis $A$ and $B$. The grey-shaded area is the primitive cell $\Sigma$. 
	The honeycomb lattice constant $a$ is defined as the distance between nearest-neighbor sites.
	}
\end{figure}

The Bravais lattice defined by \eqref{eq:avectors} is a {\it triangular} one. We opt here for the diamond-shaped primitive cell $\Sigma$ delineated by the two Bravais lattice vectors as a tiling for the optical potential \eqref{eq:honeypot}; see Fig.~\ref{fig:basis}. 
Another possible choice would have been the hexagonal Wigner-Seitz cell~\cite{ashcroft76}. This cell is useful when discussing the symmetry group of the lattice.

To proceed further one now needs to analyze the structure of the optical potential \eqref{eq:honeypot} inside the primitive cell. In passing, we mention here that red detuned ($\delta<0$) lasers give $V_0<0$ and there is only one potential minimum in each primitive cell $\Sigma$. Upon trapping atoms in these potential minima, one gets a triangular lattice that is not of graphene type. This situation is interesting in view of quantum magnetism and frustration phenomena \cite{lewenstein07} but it is not the situation we want to study here.

\subsubsection{The honeycomb structure}

\begin{figure*}[tb]
\begin{picture}(510,241)(-27.5,0)
\put(0,100){\includegraphics[bb=95 425 308 561,clip=]{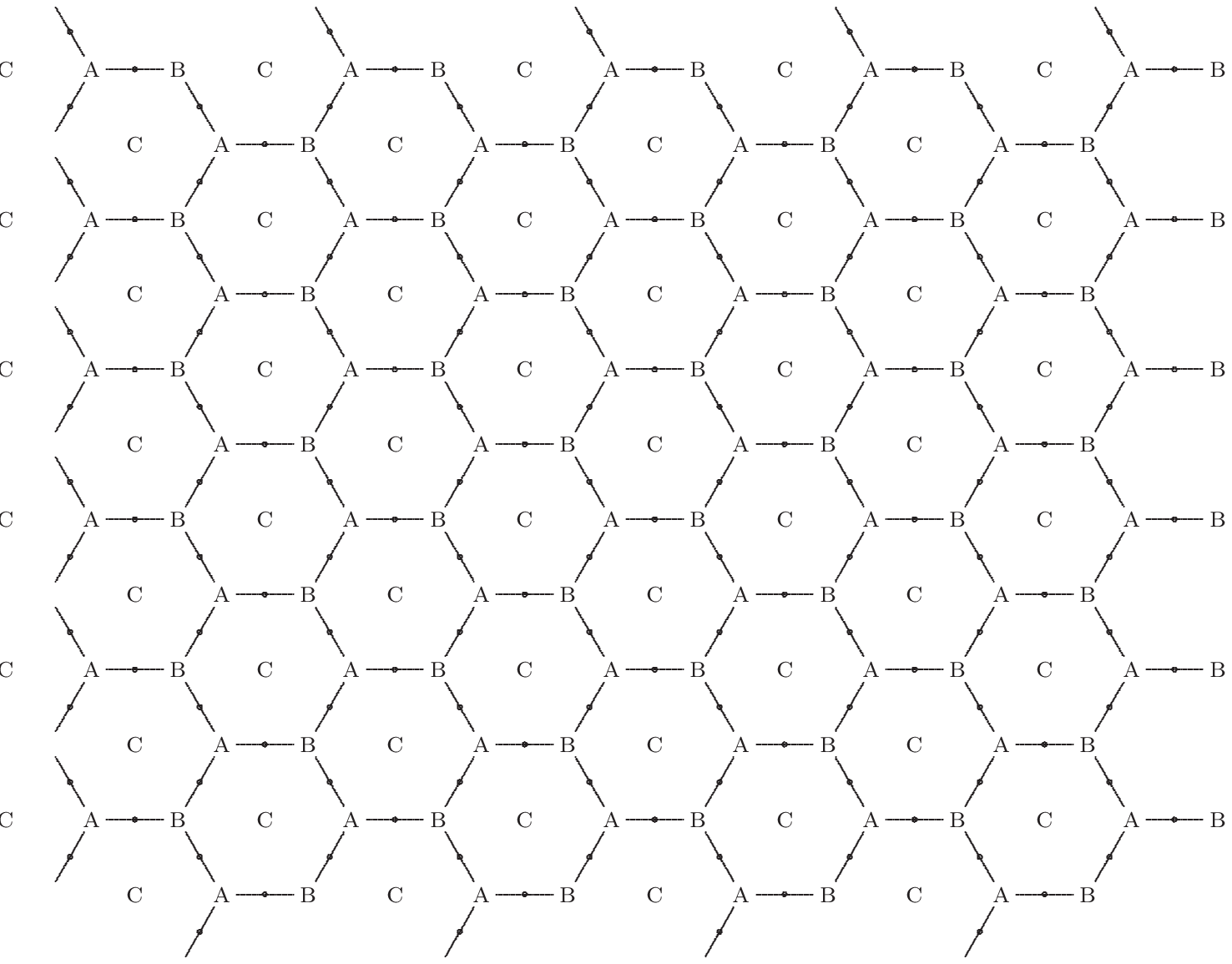}}
\put(-20,0){\includegraphics{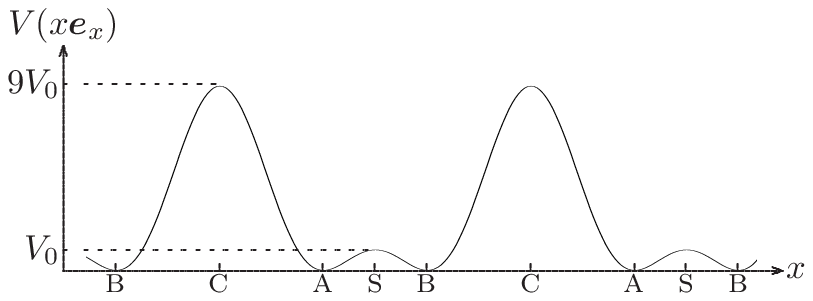}}
%
\put(240,0){\includegraphics[bb=150 470 365 707,clip=]{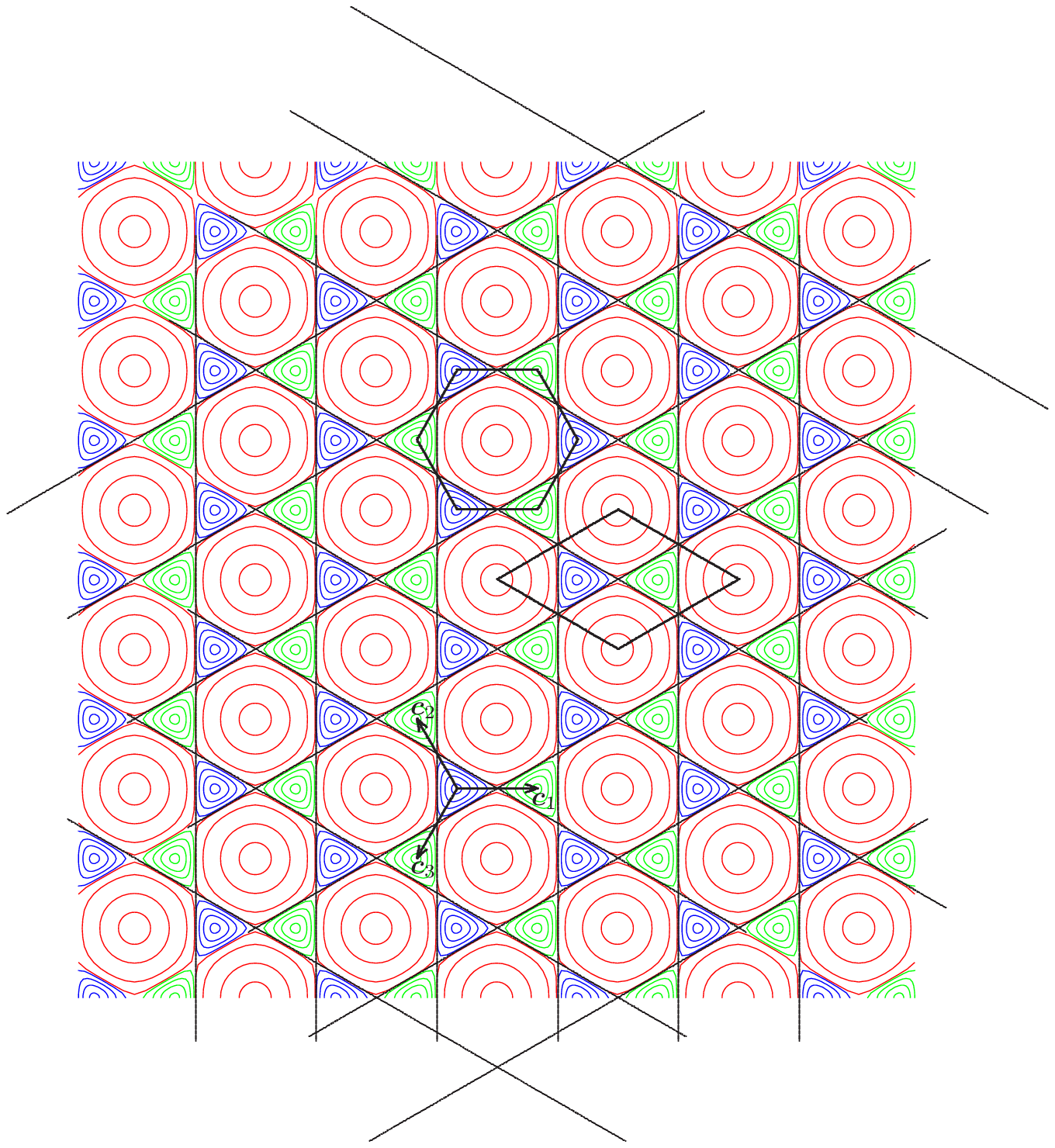}}
\end{picture}
\caption{\label{fig:ABChexagon}[Color online]
  \textbf{Left:}
  The honeycomb pattern composed of the triangular lattices of minima at sites
  \textsc{a} and \textsc{b}, of maxima at sites \textsc{c}, as well as of the
  saddle points between neighboring \textsc{a} and \textsc{b} sites (marked by
  dots).
  The bottom plot shows the potential along the $x$ axis which is one of the
  \textsc{\dots{}abcabc\dots} lines with $x=0$ at a \textsc{c} site.
  The saddle points \textsc{s} appear as local maxima here, with a height that
  is one ninth of the global maxima at sites \textsc{c}.
  Cold atoms trapped in this optical potential would be found at the
  \textsc{a} and \textsc{b} sites.
  \textbf{Right:}
  Equipotential lines for the optical honeycomb potential
  \eqref{eq:HCpotential}.
  Along the straight black lines that connect the saddle points,
  we have $V(\vec{r})=V_0$.
  The red closed circular curves fill out a hexagonal area, centered at the points of
  maximal potential; from inside out the respective values are
  $V(\vec{r})=8V_0$, $5V_0$, $2V_0$, and $1.05V_0$.
  The closed curves in blue and green fill out areas of the shape of
  equilateral triangles, their centers are the minima that constitute the
  \textsc{a} sublattice (blue) or the \textsc{b} sublattice (green); along the
  curves the potential has the values $V(\vec{r})=0.95V_0$, $0.6V_0$,
  $0.3V_0$, and $0.05V_0$.  One primitive diamond-shaped unit tile $\Sigma$ spanned by $\vec{a}_1$ and $\vec{a}_2$ is traced out. It contains two different minima, one of \textsc{a}-type (in blue, on the left inside) and one of  \textsc{b}-type (in green, on the right inside). The trine of the ${\textsc{a}\to\textsc{b}}$ displacement vectors  \eqref{eq:cvectors} is indicated as well. Finally, for completeness, we also trace out the Bravais Wigner-Seitz unit tile. It is a hexagon centered at a potential maximum and with potential minima at its corners.
  }
\end{figure*}

When the optical lattice is instead blue-detuned ($\delta > 0$), $V_0$ is positive and atoms are ``weak-field seekers''. The potential minima coincide with the minima of the electric field strength, and the maxima coincide as well. By choice of coordinate system, the maxima locate at the Bravais sites and the dimensionless potential \eqref{eq:honeypot} has its maximal value of ${v(\vec{0})=9}$ at the corners $O, P,Q,R$ of the diamond-shaped primitive cell $\Sigma$, see Fig.~\ref{fig:ABChexagon}. 

Two different potential minima, given by the zeros of the total dimensionless field amplitude $f(\bm{r})$, are found in $\Sigma$ at
\begin{equation}
  \label{eq:A+B}
  \vec{r}_\textsc{a}^{\ }=\tfrac{1}{3}(\vec{a}_1+\vec{a}_2)=\frac{\Lambda}{\sqrt{3}}\vec{e}_x\quad\mbox{and}\quad\vec{r}_\textsc{b}^{\ }= 2\vec{r}_\textsc{a}^{\ }\,,
\end{equation}
respectively. From a crystallographic point of view, $\Sigma$ is a primitive cell with a {\it two-point} basis. By applying repeated Bravais translations on $\Sigma$, one generates two different sublattices of potential minima, one made up of \textsc{a}-type sites and the other made of \textsc{b}-type sites, see Fig.~\ref{fig:basis} and Fig.~\ref{fig:ABChexagon}. Altogether the potential minima are organized in a honeycomb structure reminiscent of the positions of the carbon atoms in graphene sheets.

The three displacements that move an \textsc{a} site to
a neighboring \textsc{b} site --- they translate the \textsc{a} sublattice to
the \textsc{b} sublattice --- are parameterized by
\begin{eqnarray}
  \label{eq:cvectors}
  \vec{c}_1&=&\frac{1}{3}(\vec{a}_1+\vec{a}_2)=a\vec{e}_x\,,\nonumber\\
  \vec{c}_2&=&\frac{1}{3}(\vec{a}_2-2\vec{a}_1)
              =a\frac{-\vec{e}_x+\sqrt{3}\vec{e}_y}{2} \,,\nonumber\\
  \vec{c}_3&=&\frac{1}{3}(\vec{a}_1-2\vec{a}_2)
              =a\frac{-\vec{e}_x-\sqrt{3}\vec{e}_y}{2} \,,
\end{eqnarray}
where ${a=\biglb|\vec{c}_j\bigrb|=\Lambda/\sqrt{3}=4\pi/(3\kappa)=2\lambda_L/\sqrt{27}}$
is the honeycomb lattice constant. It is the distance from an \textsc{a} site to a neighboring \textsc{b} site,
or the distance from the center of the hexagon of minima to one of its corners.

Halfway between two neighboring minima, the potential has saddle points where
$v(\vec{r})=1$. They are located at the center and at the middle of the edges of $\Sigma$, see Fig.~\ref{fig:ABChexagon}. As the saddle points on opposite sides of $\Sigma$ are connected
by Bravais displacements, there are therefore three nonequivalent
triangular sublattices of saddle points, and we thus count three saddle points per primitive cell.

We also note that the potential is invariant under $120^\circ$ rotations around the locations of
the potential minima and maxima and, therefore, that the potential is isotropic in
the vicinity of these points. We anticipate that the local harmonic oscillator potential at a minimum will be isotropic; see \eqref{eq:LocOsc} below. By contrast, the corresponding local potential at a saddle point is not isotropic. 

All these matters are illustrated in Fig.~\ref{fig:ABChexagon},  where
we clearly identify the various triangular sublattices.
Cold fermionic atoms trapped in this optical potential would be found at the
\textsc{a} and \textsc{b} sites, similar to the binding of electrons in
graphene to the carbon ions.

As a side remark, it may be worth mentioning that the saddle points affect the
classical dynamics of a particle evolving in the honeycomb potential with a
sufficiently large energy.
Since the potential is nonseparable and angular momentum is not conserved
here, the saddle-points could be the seed for instabilities in which case the
motion could turn out to be nonintegrable and chaotic. If so, this chaotic behavior
should then be revealed, for example, in the statistical
properties of the quantum spectra, whose level spacing fluctuations is expected
to be described by the gaussian orthogonal ensemble~\cite{bohigas05}.

\subsubsection{Optical honeycomb potential and graphene}
\label{sec:HCvsGRAPH}
In graphene sheets, the electrostatic potential that governs the dynamics of electrons, the sum of the Coulomb potentials of the carbon ions, exhibits the symmetries associated to a honeycomb pattern. Of course, in the finer details, the optical dipole potential of \eqref{eq:HCpotential} and \eqref{eq:honeypot} differs markedly from the graphene potential. In particular, the very strong forces that the electrons in graphene experience close to the ions have no counterpart in the optical lattice, and the interaction between the atoms loaded into the optical potential is quite different from the electric repulsion between electrons. Nevertheless, the common symmetry group implies great similarities between the band structures of the two potentials, and in the respective parameter regimes where the tight-binding approximation is valid, the effective Hamilton operators are virtually identical. In particular, experiments made with atoms offers new knobs to play with and, with due attention to the difference between the two physical systems, these observations may deepen our understanding about phenomena observed with graphene samples.

In a very definite sense, the honeycomb potential \eqref{eq:HCpotential} is
the simplest of all graphene-type potentials \cite{followup}.
Their general form is a Fourier sum over the Brillouin vectors,
\begin{equation}
  \label{eq:graphtype}
  V(\vec{r})=\sum_{\vec{Q}\in\tilde{\mathcal{B}}}\Exp{\mri\vec{Q}\cdot\vec{r}}v_{\vec{Q}}^{\ }
\qquad\mbox{with $v_{-\vec{Q}}^{\ }=v_{\vec{Q}}^*$}
\,.
\end{equation}
The various symmetry properties of a honeycomb potential ensure that the $v_{\vec{Q}}$s are grouped into sets of closely related coefficients. If one coefficient in \eqref{eq:graphtype} is nonzero, a whole set of closely related coefficients have corresponding nonzero values as well.

Other than the trivial constant solution ${V(\vec{r})=v_{\vec{0}}^{\ }}$, the simplest case is obtained when all coefficients vanish except for the set associated with
${v_{\vec{b}_1}^{\ }=V_0}$ and, by convention, ${v_{\vec{0}}^{\ }=3V_0}$. This yields the honeycomb potential \eqref{eq:HCpotential} with $v(\vec{r})$ of \eqref{eq:honeypot}.

\section{Massless Dirac fermions}

\subsection{Band structure in the hopping picture}
In the hopping picture, one envisions the particle as hopping from site to
site with some quantum mechanical hopping (or tunneling) amplitude.
In the simplest situation, all sites have the same energy, only hops between
nearest-neighbors sites are considered and all hopping amplitudes take on the
same complex value $t_0$.
The one-particle quantum dynamics is then conveniently described using second
quantization.
In the present situation, as we have two different sub-lattices, one has to
introduce two sets of fermionic annihilation and creation operators, one for
the \textsc{a} sites, $(a^{\ }_{\vec{i}\sigma}, a^\dag_{\vec{i}\sigma})$, and one
for the \textsc{b} sites $(b^{\ }_{\vec{j}\sigma}, b^\dag_{\vec{j}\sigma})$, where
$\vec{i}$ and $\vec{j}$ label the sites in the two-dimensional lattices while
$\sigma$ stands for the spin index or any other pertinent quantum number of
the particle.
The second-quantized Hamilton operator then reads
\begin{equation}\label{eq:hop1}
H =\!\sum_{\langle \vec{i},\vec{j} \rangle,\sigma}
   \bigl(t^{\ }_0 b_{\vec{i}\sigma}^\dag a^{\ }_{\vec{j}\sigma}
         +  t_0^* a_{\vec{i}\sigma}^\dag b^{\ }_{\vec{j}\sigma}\bigr)
  + \epsilon \sum_{\vec{i}\sigma}\bigl(a_{\vec{i}\sigma}^\dag a^{\ }_{\vec{i}\sigma}
       - b_{\vec{i}\sigma}^\dag b^{\ }_{\vec{i}\sigma}\bigr)\,,
\end{equation}
where $\langle \vec{i},\vec{j} \rangle$ means that only nearest-neighbors are
included in the sum.
The model defined by this Hamilton operator accounts for hopping to
neighboring lattice sites but does not permit
a change of the internal quantum number $\sigma$ during the hop.
We have also included a possible energy mismatch $\epsilon$ between the $\textsc{a}$
and $\textsc{b}$ sites \cite{semenoff84}.
Using the Fourier transform in $\Omega$ of the fermionic operators, the
right-hand side of \eqref{eq:hop1} can be recast into the form
\begin{equation}
H = \sum_{\vec{k} \in \Omega, \sigma} (a^\dag_{\vec{k}\sigma}, b^\dag_{\vec{k}\sigma})
\left(
\begin{array}{cc}
\epsilon & z^{\ }_{\vec{k}} \\
z^*_{\vec{k}} & -\epsilon
\end{array}\right)
\left(\begin{array}{c}
a_{\vec{k}\sigma}^{\ }  \\
b_{\vec{k}\sigma}^{\ }
\end{array}\right)
\end{equation}
with
\begin{equation}
\label{eq:zk}
z_{\vec{k}} = t_0 \, \sum_n \Exp{\mri\vec{k} \cdot \vec{c}_n}\,,
\end{equation}
from which we get the band spectrum
\begin{equation}
\label{eq:bs1}
\epsilon_{\pm}(\vec{k}) = \pm \sqrt{\epsilon^2 + \biglb|z^{\ }_{\vec{k}}\bigrb|^2}\,.
\end{equation}
As expected from the fact that the honeycomb lattice consists of two distinct
sublattices, we find two bands: a conduction band ($+$) and a valence band
($-$).
These bands are here independent of the spin index $\sigma$, meaning that each
$\vec{k} \in \Omega$ accommodates ${2\sigma+1}$ internal states per subband.
Without any real loss of generality, we will stick to spin-$\frac{1}{2}$
fermions in the sequel.
As readily checked, $z_{\vec{k}}$ vanishes when
\begin{equation}
1+\Exp{\mri\vec{k}\cdot\vec{a}_1}+\Exp{\mri\vec{k}\cdot\vec{a}_2}=0\,,
\end{equation}
which is solved by the corners $K$ and $K'$ of $\Omega$ since ${\vec{K}\cdot\vec{a}_2=\vec{K}'\cdot\vec{a}_1=2\pi/3}$.
We thus see that the conduction and the valence bands are gapped by
$\epsilon$, a situation typical of a metal when the lattice is filled with
particles.
When there is exactly one particle per site (a situation known as
half-filling), all levels in the valence band are filled at zero
temperature, and the Fermi energy $E_F$ (the energy of the highest filled
level) precisely cuts the energy surface at the $K$ and $K'$ points.
In this case the low-energy excitations of the system can be described by
linearizing the band spectrum in the neighborhood of $K$ and $K'$.
Denoting by $\vec{q}= \vec{p}/\hbar$ the small displacement from either $K$ or
$K'$, the linearization of $z_{\vec{k}}$ gives
\begin{equation}
	\biglb|z_{\vec{k}}\bigrb| \approx \frac{3a\biglb|t_0\bigrb|}{2} \biglb|\vec{q}\bigrb|
                    = \hbar v_0 \biglb|\vec{q}\bigrb|
                    = \biglb|\vec{p}\bigrb|v_0\,,
\end{equation}
where the quantity $v_0 = 3a\biglb|t_0\bigrb|/(2\hbar)$ is called the
\emph{Fermi velocity} in the solid-state community.
We adopt this terminology although it is somewhat unfortunate,
because it has nothing to do with the standard Fermi velocity $\sqrt{2E_F/m}$,
which depends on the actual mass of the particle.

The dispersion relation now takes on the very suggestive form
\begin{equation}
\label{eq:bs2}
	\epsilon_{\pm}(\vec{p})\approx \pm \sqrt{m_*^2 v_0^4 + p^2v_0^2}
\end{equation}
that is typical of a relativistic dispersion relation with particle-hole
symmetry.
The effective mass $m_*$, defined through $\epsilon = m_* v_0^2$, appears thus
as the rest mass of the excitations and relates to the energy imbalance of the
two sub-lattices.
The Fermi velocity $v_0$ is the analog of the velocity of light in relativity.

The effective Hamilton operator that is derived from these considerations and
describes the dynamics of the excitations around $K$ and $K'$,
\begin{equation}
H = \int \!\frac{(d\vec{r})}{(2\pi)^2} \, \overline{\psi}(\vec{r})
\left(
\begin{array}{cc}
\mri \vec{\gamma}\cdot \vec{\nabla} + m_* & 0 \\
0 & \mri \vec{\gamma} \cdot \vec{\nabla} -  m_*
\end{array}\right)
\psi(\vec{r})\,,
\end{equation}
where $\psi(\vec{r})$ is a 4-component Dirac spinor encapsulating the
excitations around $K$ and $K'$ while $\overline{\psi}=\psi^\dagger\left(\begin{array}{cc}\gamma^0 & 0 \\
0 & \gamma^0\end{array}\right)$, generates an equation of motion that
resembles the Weyl-Dirac equation in two dimensions.
This is why the name \emph{Dirac points} is given to $K$ and $K'$
(see Refs.~\cite{semenoff84, neto09} for more details).
In this two-dimensional context, the Dirac matrices are
$\gamma^\mu = (\gamma^0, \boldsymbol{\gamma})
= (\sigma_z, \mri \sigma_x, \mri\sigma_y)$
in terms of the standard Pauli matrices.

\begin{figure}[tb]
\includegraphics[scale=1.1]{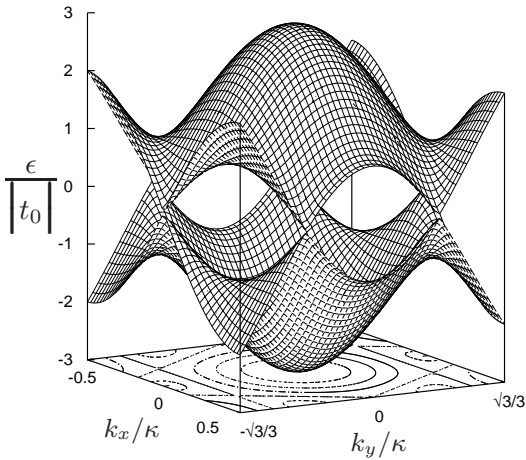}
\caption{\label{fig:honeyband}
The tight-binding band structure of graphene (in units of the tunneling
strength $\biglb|t_0\bigrb|$) as a function of $\vec{k}\in\Omega$ in units of
$\kappa=\sqrt{3}k_L$. The origin of energy has been chosen at the Dirac points
and the axis ranges are $\biglb|k_x/\kappa\bigrb| \leq 1/2$ and
$\biglb|k_y/\kappa\bigrb| \leq \sqrt{3}/3$. The bottom contour lines are lines of constant $\biglb|\epsilon\bigrb|/\biglb|t_0\bigrb|$.}
\end{figure}

\begin{figure}[tb]
\includegraphics[scale=1.0]{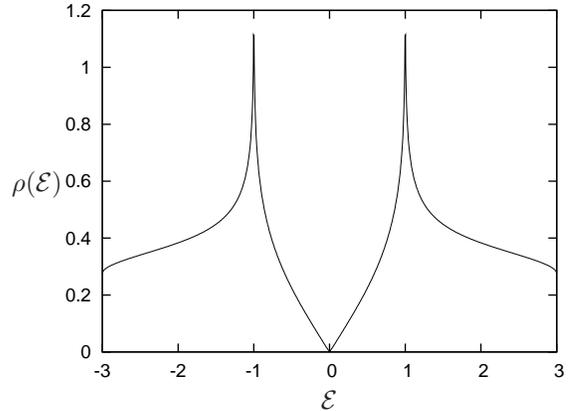}
\caption{\label{fig:doshoney}
The noninteracting density of states per unit cell and per spin component
$\rho(\mathcal{E})$ as a function of the reduced energy $\mathcal{E}=
E/\biglb|t_0\bigrb|$.
The origin of energy has been chosen at the Dirac points.
When $\mathcal{E} \ll 1$, then
$\rho(\mathcal{E}) \approx 2 \biglb|\mathcal{E}\bigrb|/(\sqrt{3}\pi)$ and the density of states
 vanishes at $\mathcal{E}=0$, a signature of the semi-metal behavior.
Note the logarithmic Van Hove singularity at $\biglb|\mathcal{E}\bigrb| = 1$.}
\end{figure}

When $\epsilon$ vanishes, as is the case in real graphene where all lattice
sites have the same energy, then $\epsilon_{\pm}(\vec{k}) = \pm \biglb|z_{\vec{k}}\bigrb|$
and the two bands are degenerate at the corners of $\Omega$ where they display
circular conical intersections (see Fig.~\ref{fig:honeyband}).
In the literature, this situation is referred to as a semi-metal or a zero-gap
semi-conductor and the corresponding low-energy excitations are known as
massless Dirac fermions.
The total band width is $W=6\biglb|t_0\bigrb|$ and, at half-filling, the Fermi
energy $E_F = 3\biglb|t_0\bigrb|$ (taking the energy origin at the lower band
minimum) precisely slices the energy bands at the Dirac points. Hence the
Fermi surface reduces to these two points, so that the density of states
vanishes there \cite{neto09}, see Fig.~\ref{fig:doshoney}.

\subsection{Tight-binding approximation}
\label{subsec:tbCalcHoneycomb}

Mindful of possible experiments, the hopping parameter $t_0$ appears to be an
important amplitude to evaluate.
We report three different methods for estimating its strength
$\biglb|t_0\bigrb|$.
We will start with the familiar tight-binding approximation using localised
Wannier functions \cite{wannier1937, wallace1947} that are further
approximated by Gaussians.
We will then develop a more accurate semi-classical calculation based on an
instanton approach \cite{milnikov2001}. We will compare both results to a
brute-force exact numerical computation.

As a consequence of Bloch's theorem \cite{ashcroft76,marder2000},
the energy spectrum of an atom of mass $m$ moving in the honeycomb lattice
potential is obtained from
\begin{equation}
\label{H-Bloch}
\mathcal{H} \psi_{n\vec{k}}(\vec{r})
= \biggl[-\frac{\hbar^2\vec{\nabla}^2}{2m}+ V(\vec{r})\biggr]
  \psi_{n\vec{k}}(\vec{r})
= \epsilon_n(\vec{k}) \psi_{n\vec{k}}(\vec{r})\,,
\end{equation}
where we dropped the spin index $\sigma$ which is not essential here.
The Bloch waves $\psi_{n\vec{k}}$ are given by
\begin{equation}
\label{u-waves}
\psi_{n\vec{k}}(\vec{r}) = \Exp{\mri\vec{k} \cdot \vec{r}} u_{n\vec{k}}(\vec{r})
\end{equation}
with $\vec{k} \in \Omega$, $n$ the band index, and $u_{n\vec{k}}(\vec{r})$ is a
$\mathcal{B}$-periodic function.
The latter can be conveniently expanded using Wannier functions
\cite{ashcroft76, marder2000, blakie04} in accordance with
\begin{equation}
u_{n\vec{k}}(\vec{r}) = \sum_{\vec{R} \in \mathcal{B}}
\Exp{-\mri\vec{k} \cdot (\vec{r}-\vec{R})} w_n(\vec{r}-\vec{R})\,.
\end{equation}
Wannier functions are very useful in describing models where particles are
localized in space, such as the Hubbard model \cite{hubbard1963}.
They form an orthonormal basis set of functions centered at different Bravais
lattice sites which are copies of the same ``seed'' functions defined in a
given primitive cell.
The localization properties of the Wannier functions crucially depend on the
analyticity properties of $u_{n\vec{k}}$ as a function of $\vec{k}$ and decay
exponentially in the simple
cases~\cite{kohn1959,cloizeaux2_1964,nenciu1983,brouder07}.

In the tight-binding approximation, the atoms are assumed to be sufficiently
deeply-localized in the different potential wells where they only populate the
lowest vibrational levels.
Vibrational states in different wells are also assumed to have small overlap:
the atomic motion is thus ``frozen'' except for the small tunneling amplitude
between neighboring wells and are then effectively confined to move in the
lowest bands of the lattice.
Since the Wannier functions display the same symmetry as the local potential
structure \cite{cloizeaux1963, cloizeaux1_1964}, the natural idea here is thus
to construct tight-binding Wannier functions from linear combinations of wave
functions deeply-localized in the two potential wells of the primitive cell (the
so-called atomic orbitals) \cite{wallace1947, reich02}.
This trial wave function exploits at best the sub-lattice structure of the
honeycomb lattice and should give good results at least for the first two
bands.

After dropping the band index $n$, this approach, reminiscent of the LCAO
method (linear combination of atomic orbitals)~\cite{ashcroft76,marder2000}, leads to the ansatz
\begin{equation}
\label{eq:ansatz}
\psi_{\vec{k}}(\vec{r}) = \alpha^{\ }_{\vec{k}}\psi^{(\textsc{a})}_{\vec{k}}(\vec{r})
                     +\beta^{\ }_{\vec{k}} \psi^{(\textsc{b})}_{\vec{k}}(\vec{r})\,,
\end{equation}
where the quasi-Bloch wavefunctions
\begin{eqnarray}\label{eqn:standardpro2}
\psi^{(\textsc{a})}_{\vec{k}}(\vec{r}) &=& \sum_\textsc{a}
\Exp{\mri\vec{k}\cdot\vec{r}_\textsc{a}}
w_\textsc{a}(\vec{r}-\vec{r}_\textsc{a})\,,\nonumber\\
\psi^{(\textsc{b})}_{\vec{k}}(\vec{r}) &=& \sum_\textsc{b}
\Exp{\mri\vec{k}\cdot\vec{r}_\textsc{b}}
w_\textsc{b}(\vec{r}-\vec{r}_\textsc{b})
\end{eqnarray}
essentially live on the type-\textsc{a} sublattice and the type-\textsc{b}
sublattice, respectively. The sublattice Wannier functions $w_\textsc{a}(\vec{r})$ and $w_\textsc{b}(\vec{r})$ are normalized to unity. In the present case, we even have $w_\textsc{b}(\vec{r})=w_\textsc{a}(-\vec{r})$ due to the reflection symmetry of the potential, ${V(-\vec{r})=V(\vec{r})}$~\cite{cloizeaux1963}. We define the on-site energies as $E_a = \langle w_a| \mathcal{H} | w_a\rangle$ ($a=\textsc{a}, \textsc{b}$) and use the parametrization  $E_\textsc{a} = E_0 +\Delta$ and $E_\textsc{b} = E_0 -\Delta$ in the following, with $E_0$ the mean on-site energy and $\Delta$ half the on-site energy difference. Most importantly, the sublattice Wannier functions are orthogonal. However, obtaining their exact expressions is a difficult task and one often resorts to simple approximations that do not have this property. This is why, in view of this very common practical situation, we will consider in the following that the Wannier functions $w_\textsc{a}(\vec{r})$ and $w_\textsc{b}(\vec{r})$ can overlap.

Plugging now the ansatz \eqref{eq:ansatz}-\eqref{eqn:standardpro2} into \eqref{H-Bloch}, and only considering coupling between nearest-neighbor lattice sites, we get the $2\times 2$ homogeneous linear system
\begin{equation}
\label{eq:linear}
\left(
\begin{array}{cc}
\Delta - E & Z^{\ }_{\vec{k}} -E R_{\vec{k}} \\
Z_{\vec{k}}^*- E R_{\vec{k}}^* & - (\Delta + E)
\end{array}
\right)
\left(
\begin{array}{c}
\alpha_{\vec{k}}^{\ } \\
\beta_{\vec{k}}^{\ }
\end{array}
\right)
= 0\,,
\end{equation}
where $E= \epsilon(\vec{k})-E_0$ and with the matrix entries
\begin{eqnarray}
\label{eq:entries}
Z_{\vec{k}} &=& \sum_a t_a  \ \Exp{\mri\vec{k} \cdot \vec{c}_a} \, , \nonumber \\
t_a &=& \langle w_\textsc{a}| (\mathcal{H} - E_0)| w_{\textsc{b}_a}\rangle \, , \nonumber \\
R_{\vec{k}} &=& \sum_a \langle w_\textsc{a} | w_{\textsc{b}_a} \rangle  
  \ \Exp{\mri\vec{k} \cdot \vec{c}_a}  \, .
\end{eqnarray}
Here $\textsc{b}_a = \textsc{a} + \vec{c}_a$ is a short-hand notation for the
three \textsc{b} sites next to the \textsc{a} site.

Several remarks are in order. First one notes that the off-diagonal matrix entries depend on the energy as soon as the sublattice Wannier functions overlap. Second, as readily checked, the hopping amplitudes $t_a$ and $E$ are independent of any energy shift in the Hamilton operator and are thus independent of any particular choice for the energy origin as one expects. Note also that the values of $E_\textsc{a}$ and of $E_\textsc{b}$ do not depend on
the particular choice for point \textsc{a} or point \textsc{b} since
$\mathcal{H}$ is $\mathcal{B}$-translation invariant.
By the same token, the values of $t_a$ and of $\langle w_\textsc{a} | w_{\textsc{b}_a}\rangle$ do not depend on the particular choice of \textsc{a}, but \textsc{b} must be one of its three nearest neighbors.

To have a nonzero solution, the $2\times 2$ determinant associated to \eqref{eq:linear} has to vanish, from which we get the band structure. When the overlaps of the sublattice Wannier functions is small, $\langle w_\textsc{a} | w_{\textsc{b}_a} \rangle \ll 1$, the band structure is very well approximated by
\begin{equation}
\epsilon_\pm(\vec{k}) \approx E_0 \pm \sqrt{\Delta^2+\biglb|Z_{\vec{k}}\bigrb|^2} \, ,
\end{equation}
a form reminiscent of \eqref{eq:bs1}. For the honeycomb lattice, for which $\mathcal{H}$ is $\mathcal{B}$-periodic
and invariant under reflections, we further have ${E_\textsc{a}=E_\textsc{b}=E_0}$ and $\Delta =0$,
which implies that the effective mass $m_*$ of the Dirac fermions is indeed zero.
As a consequence, we get the two first bands as $\epsilon_\pm(\vec{k}) = E_0 \pm \biglb|Z_{\vec{k}}\bigrb|$.
Furthermore, since $V(\vec{r})$ is also invariant under $2\pi/3$ rotations
about any lattice site \textsc{a}, all three tunneling amplitudes $t_a$ from
$\textsc{a}$ to $\textsc{b}_a$ acquire the same value and $Z_{\vec{k}}$ of
\eqref{eq:entries} turns into $z_{\vec{k}}$ of \eqref{eq:zk} with
\begin{equation}
\label{eq:hoppingconstant}
t_0 = \int (d\vec{r}) \, w_\textsc{a}^*(\vec{r})(\mathcal{H}-E_0)
w_\textsc{a}(\vec{r}-\vec{c})
\,,
\end{equation}
where $\mathcal{H}$ is the differential operator of \eqref{H-Bloch} and
$\vec{c}$ is either one of the three displacement vectors in
\eqref{eq:cvectors}.

\subsection{Harmonic approximation}

To proceed further one needs an approximation for the Wannier functions
$w_\textsc{a}$ and $w_\textsc{b}$.
One possibility is to rely on the harmonic approximation of the potential
wells around sites \textsc{a} and \textsc{b}, that is to approximate $w_\textsc{a}$ and $w_\textsc{b}$ by the corresponding harmonic ground state
wave functions. We find
\begin{eqnarray}\label{eq:LocOsc}
V(\vec{r}_a+\vec{r}) &\approx& \frac{3}{4}V_0\kappa^2\vec{r}^2
= \frac{m\omega_0^2}{2}\vec{r}^2\quad\mbox{for $a=\textsc{a},\textsc{b}$}
\nonumber \\
\mbox{with}\quad\hbar \omega_0 &=& 3 \sqrt{V_0E_R}\,,
\end{eqnarray}
where ${E_R = \hbar^2k^2_L/(2m)}$ is the recoil energy of the atom.
In terms of 
${\ell =\sqrt{\hbar/(m\omega_0)}}$,
the familiar length unit of the harmonic oscillator,
the ground state wave function is
\begin{equation}\label{eq:oscWann}
w_\textsc{a}(\vec{r}_\textsc{a}+\vec{r})=w_\textsc{b}(\vec{r}_\textsc{b}+\vec{r})
\approx
\frac{1}{\sqrt{\pi}\ell}\, \Exp{- \frac{1}{2}\vec{r}^2/\ell^2}\,.
\end{equation}
From this we get $E_\textsc{a} = E_\textsc{b} = E_0 \approx \hbar \omega_0$ and the overlap integrals are simply
\begin{equation}
\langle w_\textsc{a} | w_{\textsc{b}_a} \rangle = \exp\Biggl(- \frac{2\pi^2}{9} \sqrt{\frac{V_0}{E_R}}\Biggr) \, .
\end{equation}
Keeping in mind that ${V_0\gg E_0 \gg E_R}$ in the tight-binding regime, $\langle w_\textsc{a} | w_{\textsc{b}_a} \rangle \ll 1$ and we find from \eqref{eq:hoppingconstant}
\begin{equation}\label{eq:HOtunnel}
t_0 \approx - \biggl(\frac{\pi^2}{3}-1\biggr) \, V_0 \,
\exp\Biggl(- \frac{2\pi^2}{9} \sqrt{\frac{V_0}{E_R}}\Biggr)\,,
\end{equation}
at leading order. However, since the hopping amplitude is given by the overlap integral of the localized
wave functions $w_\textsc{a}$ and $w_\textsc{b}$ of two neighboring sites, we
see that the value of $t_0$ crucially depends on the tails of these wave
functions.
Wannier functions often decay exponentially and, therefore, they cannot be
realistically approximated by Gaussian wave functions.
Hence \eqref{eq:HOtunnel} can, at best, serve as a rough underestimate
\cite{dalibard08}.
In the next section we will derive a reliable and accurate estimate of the
tunneling amplitudes in the tight-binding regime by use of the instanton
method.

\subsection{Semiclassical estimate}

Using $k_L^{-1}$, $\sqrt{V_0/m}$, $V_0$, and $\sqrt{m/(k_L^2V_0)}$ as length,
velocity, energy, and time units, respectively, the Schr\"odinger equation can
be conveniently recast into a dimensionless form that features an effective
Planck's constant $\hbar_e$ (we keep the same symbols for the rescaled
variables for simplicity),
\begin{equation}
\mri \hbar_e \partial_t \psi = -\frac{\hbar^2_e}{2} \vec{\nabla}^2\psi
+ v(\vec{r}) \psi\,, \quad \hbar_e = \sqrt{\frac{2E_R}{V_0}}\,,
\end{equation}
with $v(\vec{r})$ given by \eqref{eq:honeypot}, here expressed in rescaled
units.
In the tight-binding approximation it is assumed that $V_0 \gg E_R$, and thus
${\hbar_e \ll 1}$.
In this situation, semiclassical methods provide very efficient and very
accurate ways for evaluating dynamical and spectral quantities of interest.
They generally amount to evaluating integrals with the aid of semiclassical
expressions for the quantum propagator, derived from its Feynman-path integral formulation
through stationary-phase approximations around the classical trajectories~\cite{gutzwiller1990}.

For example, it is well-known that the energy splitting between the two lowest
energy levels of an atom moving in a one-dimensional symmetric double well can
be accurately calculated using the WKB method~\cite{gutzwiller1990}.
This WKB method can be extended to several dimensions and in the sequel we
will derive a semiclassical estimate of $t_0$ for the honeycomb lattice using
the method proposed in \cite{milnikov2001}.
It amounts to evaluating $t_0$ using the classical complex trajectory (in
rescaled units) that connects \textsc{a} and \textsc{b} through the classically
forbidden region --- the so-called instanton trajectory.

Using $\hbar \omega_0$ as an order of magnitude for the vibrational level
inside a potential well, we see that in the rescaled units, this energy is
$\hbar \omega_0/V_0 = 3\hbar_e/\sqrt{2} \ll 1$.
So we can simply look for the instanton trajectory at zero energy.
In rescaled units, the hopping amplitude is then expressed as
\begin{equation}
\label{semi-t}
\frac{\biglb|t_0\bigrb|}{V_0}=\alpha \sqrt{\hbar_e} \Exp{-S_0/\hbar_e}\,,
\end{equation}
where $S_0$ is the (rescaled) classical action along the zero-energy instanton
trajectory, and the numerical factor $\alpha$ is obtained from integrating out
the fluctuations around the zero-energy instanton trajectory (see below).

As the zero-energy instanton fully runs in the classically forbidden region,
the variables take on complex values.
For our particular case, the good parameterization turns out to keep $\vec{r}$
real while taking $t=\mri \tau$ and $\vec{p}=-\mri \tilde{\vec{p}}$ purely
imaginary with $\tau$ and $\tilde{\vec{p}}$ real.
Hamilton's  classical equations of motion in the new variables are just
obtained from the original ones by flipping $V(\vec{r})$ to $-V(\vec{r})$.
The symmetry of the potential dictates that the zero-energy instanton
trajectory is simply the straight line connecting site \textsc{a} to
\textsc{b} (see Fig.~\ref{fig:ABChexagon}). In the following we calculate the instanton between \textsc{a} and
\textsc{a}$+\vec{c}_1$.
Integrating the equation of motions, one gets the instanton trajectory in
the rescaled form $\vec{r}_0(\tau) = k_Lax_0(\tau)\vec{e}_x$ with
\begin{equation}\label{eqn:instanton}	
\tan[\pi x_0(\tau)/3] = -\sqrt{3} \, \coth[3\sqrt{2}\tau/4]\,.
\end{equation}
The boundary conditions are ${x_0=1}$, ${\dot{x}_0=0}$ when
${\tau \to-\infty}$ and ${x_0=2}$, ${\dot{x}_0=0}$ when ${\tau \to \infty}$,
meaning that the instanton starts at \textsc{a} with zero velocity and ends at
\textsc{b} with zero velocity, the whole process requiring an infinite amount
of time.
This is indeed what is expected as both endpoints of the instanton are
instable in the reversed potential picture.
Since the energy associated with this instanton trajectory is zero, the
classical action is simply
\begin{equation}
  S_0=\int\limits_{ka}^{2ka}\! dx \; \biglb|f(x,y=0)\bigrb|
     =4\sqrt{2}{\left(1-\frac{\pi}{3\sqrt{3}}\right)} \approx 2.237\,,
\end{equation}
where $f(\vec{x,y})$ is given by \eqref{eq:fieldamp}.

The computation of $\alpha$ proves technically more demanding.
Following  \cite{milnikov2001}, it is given by the product $\alpha_1 \alpha_2$
with
\begin{eqnarray}
\alpha_1 &=& \sqrt{\frac{S_0}{2\pi}}\,
\sqrt{\frac{\det[-\partial_\tau^2+\omega_0^2]}
           {\det'[-\partial_\tau^2+\omega^2_x(\tau)]}}\,, \nonumber \\
\alpha_2 &=& \sqrt{\frac{\det[-\partial_\tau^2+\omega_0^2]}
                  {\det[-\partial_\tau^2+\omega^2_y(\tau)]}}\,.
\end{eqnarray}
Here ${\omega^2_a(\tau) = (\partial^2_a v)(\vec{r}_0)}$ ($a=x,y$) is the
curvature of the rescaled potential along the zero-energy instanton trajectory
$\vec{r}_0(\tau)$ while $\omega_0$ is the curvature of the rescaled harmonic
potential approximation around \textsc{a}; see \eqref{eq:LocOsc}.
In rescaled units, we have $\omega_0 = 3/\sqrt{2}$.
The prime in the formula for $\alpha_1$ means that the determinant is
calculated by excluding the eigenspace of the operator
${-\partial_\tau^2+\omega^2_x}$ with the smallest eigenvalue.

The determinants of the differential operators involved in the computation of
$\alpha$ stem from the linear stability analysis of the dynamical flow in the
neighborhood of the zero-energy instanton trajectory as encapsulated in the
monodromy matrix.
They can be straightforwardly computed from solutions of the linear
Jacobi-Hill equations of degree 2 associated with these differential operators
\cite{schulman1981}.
For example, $\alpha_2$ is solved as
\begin{equation}
\alpha_2 = \lim_{T\to\infty} \sqrt{\frac{J_0(T)}{J(T)}}
\end{equation}
where the Jacobi fields $J(\tau)$ and $J_0(\tau)$ satisfy the differential
equations
\begin{eqnarray}\label{eqn:jacobieq}
\frac{d^2J(\tau)}{d\tau^2}-\omega^2_y(\tau)J(\tau) &=& 0\,,\nonumber\\
\frac{d^2J_0(\tau)}{d\tau^2}-\omega^2_0J_0(\tau)&=&0\,,
\end{eqnarray}
with initial conditions
\begin{eqnarray}
J_0(-T) &=& J(-T) = 0\,,\nonumber\\
\dot{J}_0(-T) &=& \dot{J}(-T) = 1\,.
\end{eqnarray}
The interested reader is referred to \cite{milnikov2001, schulman1981} for
details.
We simply give here the final result for the honeycomb lattice:
\begin{equation}
\alpha_1 = \sqrt{\frac{27\sqrt{2}}{\pi}} \approx 3.486\,, \quad
\alpha_2 \approx 0.449\,, \quad \alpha \approx 1.565\,.
\end{equation}
Recasting the semiclassical calculation of the tunneling amplitude in
units of the recoil energy finally yields
\begin{equation}
\label{t0ER}
\frac{\biglb|t_0\bigrb|}{E_R} \approx 1.861
\left(\frac{V_0}{E_R}\right)^{\!3/4} \,
\exp\Biggl[- 1.582 \, \sqrt{\frac{V_0}{E_R}}\Biggr]\,.
\end{equation}
The same type of scaling laws has been obtained in the case of the
two-dimensional square optical lattice \cite{zwerger03, dalibard08}.
In the square-lattice geometry, however, the potential is separable and the
semiclassical calculation proves much simpler as it reduces to using the
well-known Mathieu equation.

\subsection{Numerical computation of the band structure}

\begin{figure}[tb]
\includegraphics[scale=1.0]{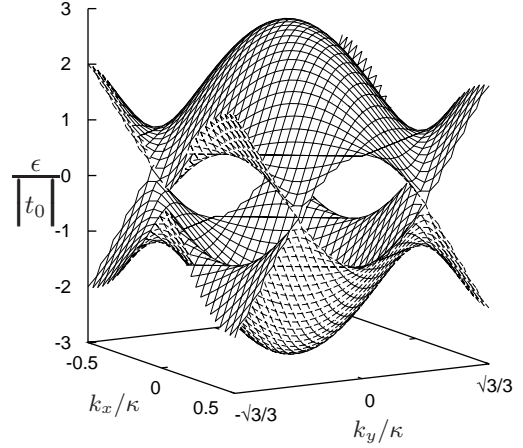}
\caption{\label{fig:bandh0.25}Numerically calculated band structure of the two lowest energy bands for $\hbar_e=0.25$ at discrete points in the Brillouin zone $\Omega$. The same conventions as in Fig.~\ref{fig:honeyband} are adopted. The value of $\biglb|t_0\bigrb|$ is determined by requiring that $\epsilon_\pm = \pm 3\biglb|t_0\bigrb|$ at the center $\Gamma$ of the Brillouin zone. The similarity with Fig.~\ref{fig:honeyband} shows that at $V_0 = 32E_R$ the tight-binding regime has already been reached.}
\end{figure}
Using Bloch's theorem and plugging \eqref{u-waves} into \eqref{H-Bloch}, we
get a family of partial differential equations for the $u_{n\vec{k}}$s labeled
by the Bloch vector $\vec{k} \in \Omega$.
After scaling variables with the same units as in the previous paragraph, the band
structure is then extracted by numerically solving
\begin{eqnarray}
\mathcal{H}_{\vec{k}} u_{n\vec{k}} (\vec{r})& =&
\epsilon_{n\vec{k}}  u_{n\vec{k}} (\vec{r})\,, \nonumber \\
\mathcal{H}_{\vec{k}} &=&\displaystyle
-\frac{\hbar^2_e}{2} (-\mri \boldsymbol{\nabla} + \vec{k})^2 + v(\vec{r})
\end{eqnarray}
for each $\vec{k} \in \Omega$ (expressed now in units of $k_L$).

\begin{figure}[tb]
	\includegraphics[scale=0.6]{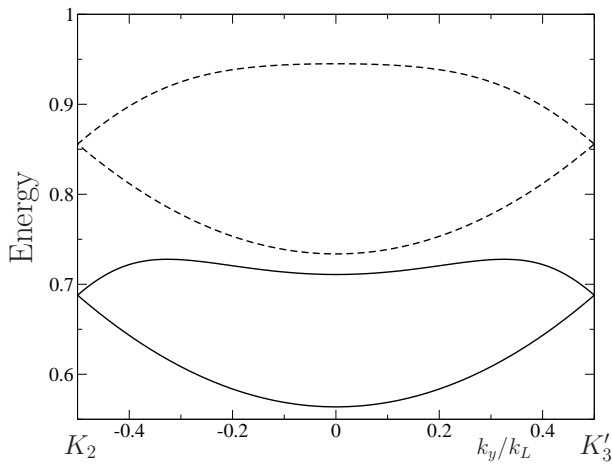}
\caption{\label{fig:wannierTransition}
Band structure for nearly-free particles moving in a weak
honeycomb optical potential in units of $V_0$.
The first $2$ levels are plotted as a function of $k_y/k_L$ at $k_x/k_L =
\sqrt{3}/2$, so along the vertical edge of $\Omega$ from $K_2$ to $K'_3$, see
Fig.~\ref{fig:honeycombRecipLattice}.
The solid curves are obtained for $\hbar_e=\sqrt{2E_R/V_0}=\sqrt{10}$ and the
dashed ones for $\hbar_e=\sqrt{5}$.
As one can see the band structure is rather flat in the band centre but the
levels curvature increases when $\hbar_e$ is decreased.
The Dirac degeneracies in the ground state obtained at the Brillouin zone
corners are generic and can be inferred from group-theoretic
considerations. Note however that the conical intersections do not extend much
over the first Brillouin zone when the potential is weak but start to spread
when $\hbar_e$ is decreased.}
\end{figure}

The $u_{n \vec{k}}$s being $\mathcal{B}$-periodic, they are conveniently
Fourier expanded in the reciprocal lattice $\tilde{\mathcal{B}}$ according to
\begin{equation}
u_{n\vec{k}}({\bf r}) =
   \sum_{\vec{Q}\in \tilde{\mathcal{B}}} C_{n\vec{Q}} \, e^{\mri\vec{Q}\cdot{\bf r}}.
\end{equation}
The matrix representation of $\mathcal{H}_{\vec{k}}$ is sparse and banded.
It is then appropriately truncated and diagonalized such that only a small
number of coefficients $C_{n\vec{Q}}$ are actually significant for the
corresponding energy bands.
The energy bands obtained in this way are exact and one can investigate their
dependance on $\hbar_e$ as done in Fig.~\ref{fig:bandh0.25} and
Fig.~\ref{fig:wannierTransition}.

The essential feature is to realize that the band degeneracies at points $K$
and $K'$ are generic and do not depend on the actual value of the effective
Planck's constant.
Indeed the existence of two degeneracy points in the first Brillouin zone for
the honeycomb lattice is a general consequence of the lattice symmetries
\cite{lomer55, slonczewski1958}.
The lattice symmetries are encapsulated in the point group of the lattice
which is the set of operations that leave fixed one particular point of the
lattice.
The corresponding elements are rotations, reflections, inversions and their
combinations.
Combined with $\mathcal{B}$-translations, one gets the space-group of the
lattice.
The graphene point group has been analyzed by Lomer \cite{lomer55} and
contains twelve elements.
In terms of Bloch wave functions $\psi_{n{\bm k}}$, the lattice space-group
operations translate into point group operations on $\vec{k} \in \Omega$,
possibly followed by a reciprocal lattice translation to bring back the
resulting new wave vector in $\Omega$.
The key point is that degeneracies can only occur at Bloch wave vectors which
are invariant (up to reciprocal lattice translations) under the action of a
nonabelian subgroup $G$ of the point group. For the graphene this happens at
the Dirac points.
For example, at corner $K_1$, beside unity, $G$ is made of two rotations of
angles $\pm 2\pi/3$ about the centre $\Gamma$ of the Brillouin zone and three
reflexions about the lines connecting $\Gamma$ to the three points labeled
$K$.
This group of order six admits an irreducible two-dimensional representation
which explains the band spectrum degeneracy at the Dirac points.

This can be nicely illustrated in the weak $V_0$ limit (or equivalently when
$\hbar_e$ is large).
In this case, the particles are quasi-free and the band spectrum can be
understood in two steps.
First, one folds the parabolic dispersion relation of the free particle into
the first Brillouin zone (repeated-zone scheme \cite{ashcroft76}) and then one
couples crossing levels at Bragg planes by the weak potential.
At $K_1$, three plane waves fold with the same kinetic energy, namely
$\vec{K}_1=\vec{k}_1$, $\vec{K}_2=\vec{k}_2$ and
$\vec{K}_3=\vec{k}_3$ (see Fig.~\ref{fig:honeycombRecipLattice}).
The weak periodic potential then couples these three plane wave states and the
coupling matrix elements are all identical.
The eigenstates of this $3\times3$ matrix split into a singlet and a
doublet. When $V_0$ is negative, the singlet is the ground state which is
consistent with the triangular Bravais lattice obtained in this case ($\delta
<0$).
When $V_0$ is positive ($\delta >0$), the doublet becomes the ground state and
features the tip of the conical intersection between the two sub-bands when
the quasi-momentum is moved away from $K$,
see Fig.~\ref{fig:wannierTransition}.

\begin{figure}[tb]
\includegraphics[scale=1.2]{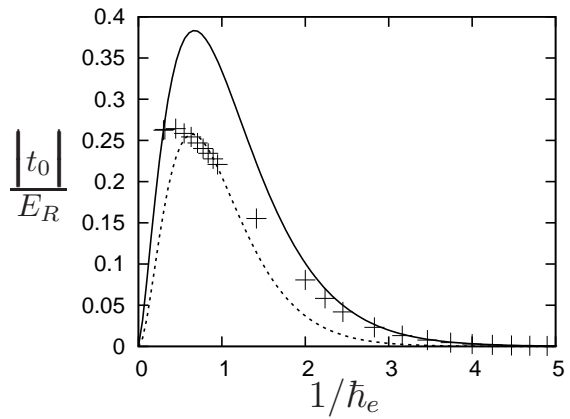}\caption{\label{fig:t0}
The hopping parameter $\biglb|t_0\bigrb|$ in units of the recoil energy
$E_R$ (crosses) as a function of the inverse of the effective Planck's
constant $\hbar_e = \sqrt{2E_R/V_0}$ as obtained from the exact numerical
computation.
The harmonic approximation (dashed curve) and the semiclassical calculation (solid
curve) of the hopping parameter have been added for comparison even if their
range of validity is restricted to the tight-binding regime $\hbar_e \ll 1$.
}
\end{figure}

\begin{figure}[tb]
\includegraphics[scale=1.2]{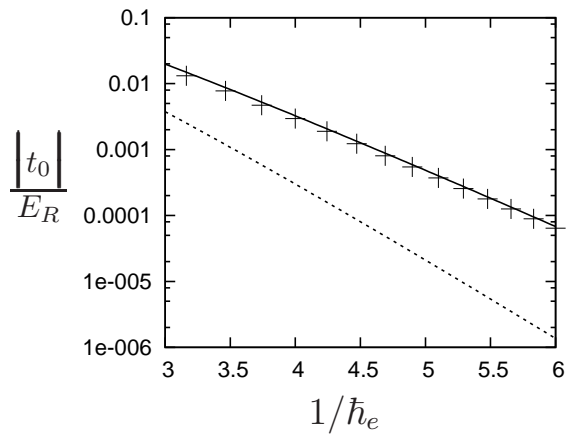}\caption{\label{fig:t0Enlarge}
The hopping parameter $\biglb|t_0\bigrb|$ (in units of the recoil energy
$E_R$) as a function of the inverse effective Planck's constant $\hbar_e =
\sqrt{2E_R/V_0}$ in the tight-binding regime where $\hbar_e \ll 1$.
As one can see, the harmonic approximation (dashed curve) is completely off.
For example at $V_0 = 32 E_R$ (or $\hbar_e = 0.25$) $\biglb|t_0\bigrb|$ is underestimated
by a factor 10 and the discrepancy gets worse as $V_0$ increases.
On the other hand, the agreement between the semiclassical calculation (solid
curve) and the exact numerical computation (crosses) just proves
excellent.}
\end{figure}

From the exact numerical calculation, one can extract the slope of the
dispersion relation at the Dirac points and then the corresponding tunneling
strength $\biglb|t_0\bigrb|$ as a function of $\hbar_e^{-1}$, see Fig.~\ref{fig:t0}.
Figure~\ref{fig:t0Enlarge} gives the comparison between the exact calculation,
the harmonic and the semiclassical calculations as a function of
$\hbar_e^{-1}$ in the tight-binding regime where $\hbar_e \ll 1$.
As one can see, the harmonic approximation is way off whereas the
semiclassical estimate proves excellent.

\subsection{Reaching the massless Dirac fermions regime}
\begin{figure}[tb]
\includegraphics[scale=1.0]{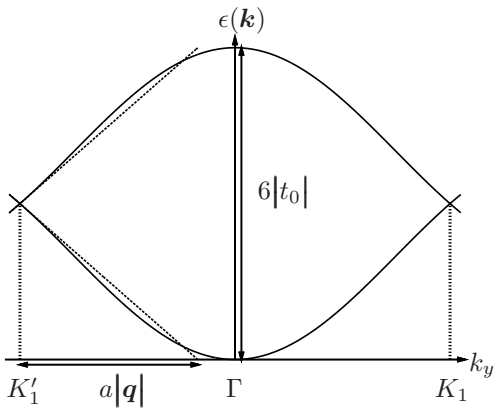}\caption{\label{fig:temp}
Cut of the linear dispersion approximation along $Oy$ at $k_x=0$ in the first
Brillouin zone $\Omega$ as compared to the actual band spectrum  in the
tight-binding regime.
At half-filling, the Fermi energy cuts the band spectrum at the Dirac points
$K$ and $K'$.
Doping the system away from half-filling moves the Fermi energy up or down but
the system can still be described in terms of massless Dirac fermions provided
$a\biglb|\vec{q}\bigrb| \ll 2$, i.e. provided the change in the Fermi energy is much less
than the band-width $W=6\biglb|t_0\bigrb|$ itself.
By the same token, thermal excitations of the system can still be described as
thermal massless Dirac fermions provided $k_BT \ll W$. }
\end{figure}

To access the massless Dirac fermions regime one first needs to completely
fill the ground state band alone, a situation known as half-filling.
This is achieved by having exactly one fermion with spin state $\sigma=\pm1/2$
per Bravais cell. Starting from a spin-unpolarized cloud of fermions,
half-filling is thus reached by loading the optical honeycomb lattice with
exactly $2$ fermions per Bravais cell, corresponding to an average number
density $\rho=1$ in the tight-binding picture.
When this is achieved, the Fermi energy slices the band structure at the Dirac
points.
For experiments that study transport phenomena, one would also need to subsequently
dope the sample away from half-filling such that the Fermi energy of the
system is varied in the linear part of the band structure.

In a usual experiment, atoms are generally held in an external harmonic
potential and the optical lattice potential is superimposed.
Reaching half-filling could then be done in two steps, first by significantly
increasing repulsive interactions $U$ between fermions through a Feshbach
resonance and then by driving the system into the Mott-Hubbard phase with one
fermion per site as done in \cite{jordens08, schneider08}.
Then setting $U$ to zero again should maintain the system at number density
$\rho = 1$.
Obvious candidates for such experiments are Potassium-40 as well as Lithium-6
atoms \cite{jordens08, schneider08, bourdel04}.
In the external trap, the Mott insulator appears first where the local filling
is approximately one atom per site and one needs to ensure that adding more
atoms (or increasing the chemical potential $\mu$) does not favor the
appearance of the doubly-occupied Mott phase.
This will be the case for very strong repulsion $U \gg \mu, t_0, k_BT$ in
which case one expects the entire centre of the trap to contain a Mott
insulating phase with single occupancy and negligible thermally-activated
doubly-occupied sites.
In the case of the honeycomb lattice, starting from a spin-unpolarized sample,
it is known that half-filling is reached for $U_c \sim 5 t_0$, the atoms
displaying at the same time an anti-ferromagnetic order \cite{paiva05}.
Note that $U_c \sim W$, where $W=2E_F=6\biglb|t_0\bigrb|$ is the band-width.

Doping the system could be done in the following way.
The external harmonic confinement (with angular trap frequency $\Omega_t$)
defines a characteristic length $\zeta = \sqrt{2\biglb|t_0\bigrb|/(m\Omega_t)}$ over which
the energy is shifted by precisely the tunneling energy $t_0$
\cite{rigol03,rigol04,kohl05}.
This length defines the distance over which one given lattice site is coupled
by tunneling to its neighbors.
In turn, having loaded $N_F$ fermions into the trap, one can define a
characteristic filling factor through $\tilde{\rho} = N_F
(a/\zeta)^2$. Varying $\biglb|t_0\bigrb|$ by changing the lattice potential height $V_0$
or tightening/loosening the trap by changing $\Omega_t$ would thus allow to
tune $\tilde{\rho}$ in a controlled way and hence to dope the system.

For the conical intersection at the Dirac points to significantly spread over
the Brillouin zone $\Omega$, one needs to reach the tight-binding regime where
$V_0$ is large enough (typically $V_0 > 10 E_R$ will do).
Inspection of the Taylor expansion of \eqref{eq:zk} then shows that it is
sufficient to have $\biglb|\vec{q}\bigrb|a \ll 2$ ($\vec{q}$ being the small displacement
from a Dirac point) for the band structure to be well approximated by a linear
dispersion relation around the Dirac points.
The available energy range $\Delta E$ is thus set by the band-width itself,
namely $\Delta E \ll W$. So tuning the filling factor away from half-filling
and residual thermal fluctuations will keep the system in the massless Dirac
fermions regime provided $\mu, k_BT \ll W$ (Fig.~\ref{fig:temp}).
For example, at $V_0 = 32 E_R$, the temperature constraint, as derived from
\eqref{t0ER}, is $T < T_R/50$ whereas it is $T < T_R/2$ at $V_0 = 10 E_R$.
There is thus room left for reaching the massless Dirac fermions regime within the
current state-of-art cooling technology.

\section{Robustness of the massless Dirac fermions}

As the very existence of the massless Dirac fermions regime rests on the two
conical degeneracies in the band structure, one may wonder if this regime
would resist imperfections of the system.
Indeed the argument we gave to explain the conical degeneracies relied on
group-theoretic arguments which were specific to the hexagonal symmetry of the
honeycomb structure.
In practice, it is impossible to control the laser configuration to the point
where all intensities and alignment angles would all be exactly equal.
Such imperfections in the system would obviously break the hexagonal symmetry
and one could think that the Dirac fermions would just be destroyed.
In fact, as we will see shortly, massless Dirac fermions are quite robust and survive small imperfections that are easily within
experimental reach.

\subsection{Imbalanced hopping amplitudes}

\begin{figure}[tb]
\centering
	\includegraphics[scale=1.0]{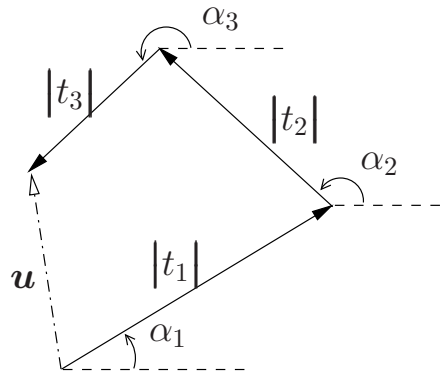}\caption{\label{fig:h12diagram}
The condition $Z_{\vec{k}} = 0$ is equivalent to cancel the resultant vector
$\vec{u}$ of three vectors, each with length $\biglb|t_n\bigrb|$ and polar angle $\alpha_n
= \vec{k}\cdot\vec{c}_n + \arg(t_n)$.
There will always be a solution provided one of the norm inequalities {$\biglb|\biglb|t_2\bigrb|
-\biglb|t_3\bigrb|\bigrb| \leq \biglb|t_1\bigrb| \leq \biglb|t_2\bigrb|+\biglb|t_3\bigrb|$} (and cyclic permutations) is satisfied.}
\end{figure}

To understand why massless Dirac fermions are robust, we will start by
analyzing the case of imbalanced hopping amplitudes as done in
\cite{hasegawa2006}.
For real graphene, this would correspond to stretching the graphene sheet.
In this case, the tight-binding band structure is given by
$\epsilon_\pm(\vec{k}) = \pm \biglb|Z_{\vec{k}}\bigrb|$, where $Z_{\vec{k}}$ is defined in
\eqref{eq:zk}.
The degeneracies are found at points $\vec{k}_D\in\Omega$ canceling
$Z_{\vec{k}} = 0$.
This condition boils down to sum up three vectors to zero in the
two-dimensional plane, see Fig.~\ref{fig:h12diagram}.
As such, a solution is only possible provided the hopping amplitudes satisfy
one of the norm inequalities given by
\begin{equation}
\label{t-condition}
\biglb|\biglb|t_2\bigrb| -\biglb|t_3\bigrb|\bigrb| \leq \biglb|t_1\bigrb| \leq \biglb|t_2\bigrb|+\biglb|t_3\bigrb|
\end{equation}
and cyclic permutations.
If this is the case, defining the angles $\varphi_{1,2} = \arg t_{2,3} - \arg
t_1$, the Dirac points solve
\begin{eqnarray}
	\cos(\vec{k}_D\! \cdot \! \vec{a}_1-\varphi_1)
&=& \frac{\biglb|t_3\bigrb|^2-\biglb|t_2\bigrb|^2-\bigrb|t_1\bigrb|^2}{2\biglb|t_1 t_2\bigrb|},\nonumber\\
	\cos(\vec{k}_D\! \cdot\! \vec{a}_2-\varphi_2)
&=& \frac{\biglb|t_2\bigrb|^2-\biglb|t_3\bigrb|^2-\biglb|t_1\bigrb|^2}{2\biglb|t_1 t_3\bigrb|},
\end{eqnarray}
subject to the condition
\begin{equation}
\biglb|t_2\bigrb| \sin(\vec{k}_D\! \cdot\! \vec{a}_1-\varphi_1)
+ \biglb|t_3\bigrb| \sin(\vec{k}_D\! \cdot \! \vec{a}_2-\varphi_2) = 0.
\end{equation}
We find the important result that the system self-adapts to changes in the
hopping amplitudes by shifting the Dirac points away from the corners of the
Brillouin zone until the norm inequalities \eqref{t-condition} break and
degeneracies disappear.
Thus, provided the hopping imbalance is not too strong, the massless Dirac
fermions do survive imperfections in the system and the hexagonal symmetry
breaking.

\begin{figure}[tb]
\centering
\includegraphics[scale=1.0]{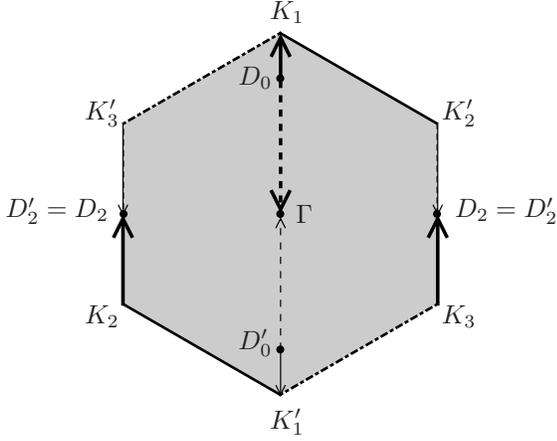}\caption{\label{fig:diracshift}
When the three hopping amplitudes $t_n$ are unbalanced, the
Dirac points are shifted in the Brillouin zone $\Omega$ and disappear when the
norm inequality $\biglb|\biglb|t_2\bigrb| -\biglb|t_3\bigrb|\bigrb| \leq \biglb|t_1\bigrb| \leq \biglb|t_2\bigrb|+\biglb|t_3\bigrb|$ is no longer
satisfied.
We depict here how the Dirac points $D_\gamma$ and $D'_\gamma$ move in
$\Omega$ when only one hopping amplitude is imbalanced, namely $t_1 \!=\!
\gamma t_0$ and $t_2\!=\!t_3\!=\!t_0$.
Points $D_\gamma$ (thick path) and $D'_\gamma$ (thin path) move along opposite
paths.
Increasing $\gamma$ from $0$, point $D_\gamma$ starts at $D_0$ and moves
upward.
It reaches point $K_1$ at $\gamma=1$ (balanced amplitudes case) then moves
along the vertical edge of $\Omega$ where it reaches its middle point $D_2$ at
$\gamma=2$.
The Dirac points cease to exist when $\gamma > 2$.
For negative $\gamma$, $D_\gamma$ moves downward from $D_0$ (dotted thick path),
reaches the zone center $\Gamma$ for $\gamma=-2$ and then ceases to exist for
$\gamma < -2$. }
\end{figure}

We illustrate this important feature in the simple case of only one imbalanced
hopping amplitude, namely $t_1 = \gamma t_0$, $t_2=t_3=t_0$.
We further choose $\gamma$ real and $0< \biglb|\gamma\bigrb| \leq 2$ for the Dirac points
to exist.
We then find two Dirac points $D_\gamma$ and $D'_\gamma$ given by $\vec{k}_D =
-\vec{k'}_D = \varphi_0 (\vec{b}_2 -\vec{b}_1)$ where $\varphi_0 \in [0,1/2]$
solves $\cos(2\pi\varphi_0) = -\gamma/2$.
This means that the two Dirac points $D_\gamma$ and $D'_\gamma$ move along
opposite paths in the Brillouin zone $\Omega$. The fact that Dirac points always come in by pairs of opposite location in $\Omega$ is generic \cite{montambaux09}.
When $\gamma$ is increased from $0$ to $2$, $D_\gamma$ starts at $\vec{k}_0 =
(3k_L/4) \, {\bf e}_y$ for $\gamma =0$, then moves along axis $Oy$ and reach
corner $K_1$ at $\gamma=1$.
Note that when $\gamma \to 0$, the physical situation is that of weakly
coupled ``zig-zag'' linear chains.
For $\gamma >1$, $D_\gamma$ leaves $\Omega$ but a translation in reciprocal
lattice brings it back on the vertical edges of $\Omega$ (technically we get
two copies of the same point).
$D_\gamma$ reaches the middle of the vertical edge at $\gamma =2$ where it
merges with $D'_\gamma$ into a {\it single} Dirac point, see
Fig.~\ref{fig:diracshift}.
Interesting physics occurs at $\gamma=2$ in connection with the quantum Hall
effect \cite{dietl08, goerbig08}.
As soon as $\gamma >2$, the degeneracy is lifted and the massless Dirac
fermions do not exist anymore.
For negative $\gamma$, $D_\gamma$ and $D'_\gamma$ move back from $\pm(3k_L/4)
\, {\bf e}_y$ to the centre $\Gamma$ of the Brillouin zone where they merge
and disappear, see Fig.~\ref{fig:diracshift}. The fact that Dirac points can only merge at the centre and mid-edge points of $\Omega$ is also generic \cite{montambaux09}.

Hence, far from being a nuisance, we see that controlling the hopping
amplitude imbalance proves an interesting way of exploring the massless Dirac
fermions physics under different circumstances by moving around the Dirac
points in the Brillouin zone.

\subsection{Optical lattice distortions}

\begin{figure}[tb]
\includegraphics[scale=0.8]{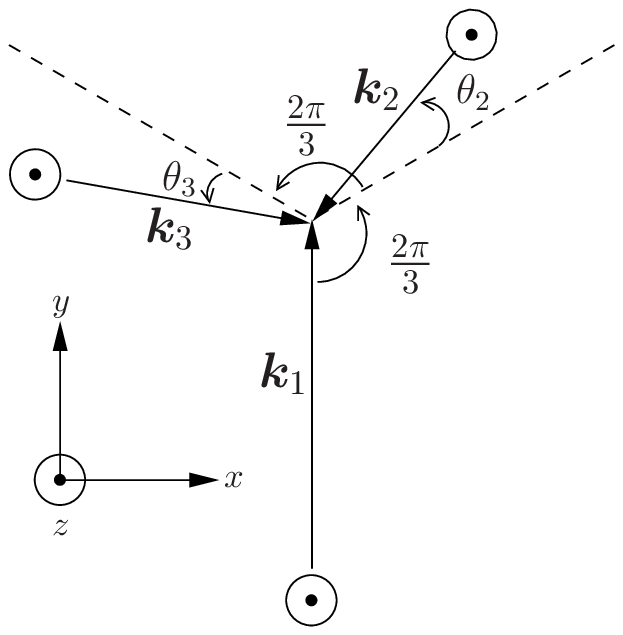}\\(a)\\
\includegraphics[scale=0.7]{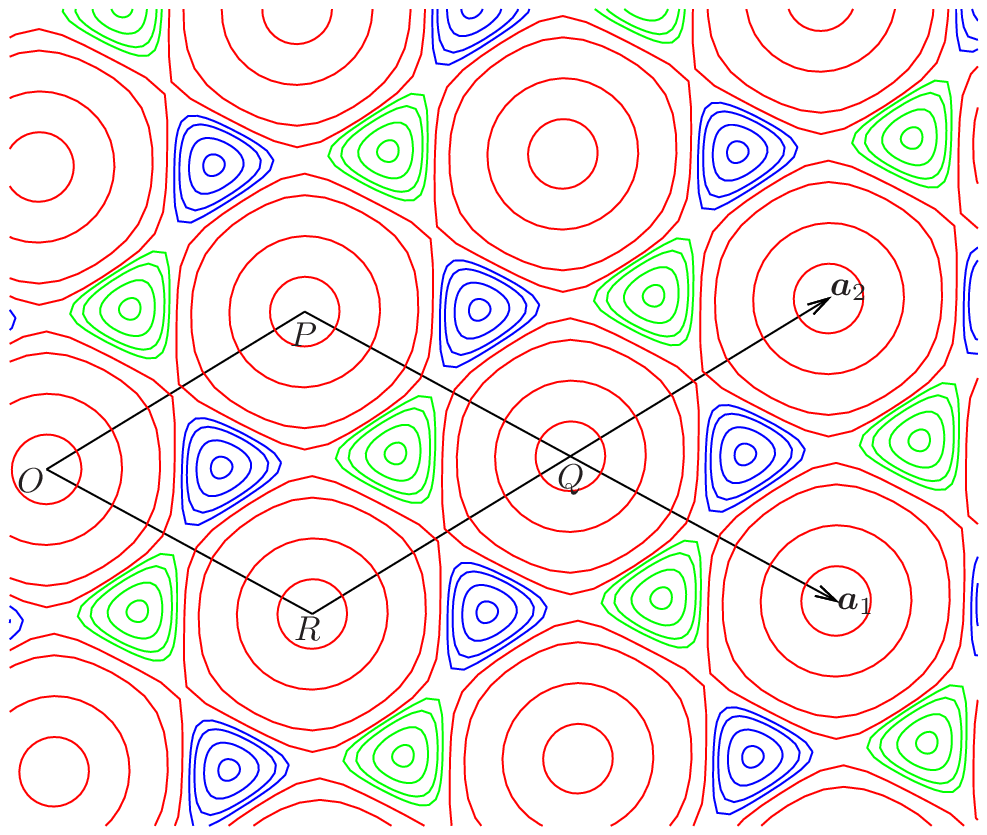}\\(b)\\
\caption{\label{fig:honeycombLatticeAsym}(a) [Color online] (a)  The
  asymmetric in-plane 3-beam configuration.
Three monochromatic and linearly-polarized laser beams with wave vectors
$\vec{k}_n$ interfere with different strengths $E_n =s_n E_0$ ($n=1,2,3$).
The respective angles depart from $2\pi/3$.
(b) Distorted optical lattice obtained with $\vartheta_3=\vartheta_2= 5 \times
10^{-2}$ and $s_1 = 1$, $s_2=1.03$, $s_3=0.97$.
For weak enough distortions, the primitive diamond-shape cell $\Sigma$ still contains two
field minima as evidenced in the plot. 
}
\end{figure}

The previous discussion concentrated on the impact of imbalanced hopping
amplitudes irrespective of the change of symmetry of the lattice potential.
We will now analyze these lattice distortions in more detail and give
quantitative estimates about the experimental degree of control which is
required to target the massless Dirac fermions regime.
We will consider in-plane laser beams with different (positive) strengths $E_n
= s_n E_0$ and with respective angles away from $2\pi/3$, see
Fig.~\ref{fig:honeycombLatticeAsym}.
It is important to note that we will always stick to imperfections which are
compatible with a two-point Bravais cell.
They will only induce distortions of the hexagonal spatial structure of field
minima but without breaking this pattern.

The new optical lattice potential is now given by $V'(\vec{r}) = V_0 \,
\biglb|f'(\vec{r})\bigrb|^2$ with the new total dimensionless field amplitude
\begin{equation}
\label{newfield}
f'(\vec{r}) = s_1 + s_2 \exp(-\mri \vec{b'}_1\cdot\vec{r}) + s_3 \exp(\mri
\vec{b'}_2\cdot\vec{r}).
\end{equation}
Here the $\vec{b'}_n$ (n=1,2) feature the new reciprocal lattice basis vectors.
They define in turn a new set of Bravais lattice basis vectors $\vec{a'}_n$
giving rise to a new primitive diamond-shaped cell $\Sigma'$. Unless the angle
mismatches vanish, the new Bravais and reciprocal lattices are no longer
hexagonal but oblique with no special symmetry except for inversion.
As a consequence, the new first Brillouin zone $\Omega'$ is still a hexagon
but {\it no longer a regular one}.

Since we assume a two-point primitive cell, the minima of the new optical
potential still identify with zeros of $f'(\vec{r})$.
Similarly with the case of imbalanced hopping amplitudes, we find two
solutions if and only if the field strengths $s_n$ satisfy one of the norm
inequalities {$\biglb|s_2-s_3\bigrb| \leq s_1 \leq s_2+s_3$} (and cyclic ones).
In this case the minima are given by
\begin{eqnarray}
	\cos(\vec{b'}_1\cdot\vec{r}) &=& \frac{s_3^2-s_2^2-s_1^2}{2s_1 s_2},
\nonumber\\
	\cos(\vec{b'}_2\cdot\vec{r}) &=& \frac{s_2^2-s_3^2-s_1^2}{2s_1 s_3},
\end{eqnarray}
subject to the condition
$s_2 \sin(\vec{b'}_1\cdot\vec{r}) = s_3 \sin(\vec{b'}_2\cdot\vec{r})$.

In the following we will examine separately the effect of strength imbalance
and angle mismatch.

\subsubsection{Critical field strength imbalance}

To give an estimate of the critical field strength imbalance beyond which the
Dirac points cannot survive, we consider the simple case of only one
imbalanced laser beam and no angle mismatch, namely $\theta_2=\theta_3 =0$,
$s_1=1+\eta$ and $s_2=s_3=1$.
In this case the Bravais lattice, the reciprocal lattice, the primitive cell
$\Sigma$ and the Brillouin zone $\Omega$ are {\it not} modified.
The new optical potential $V'(\vec{r}) = V_0 v'(\vec{r})$ reads
\begin{eqnarray}
\label{new-v}
v'(\vec{r}) &=& v(\vec{r}) + 2 \eta \,  \delta v(\vec{r}) + \eta(\eta+2),
\nonumber\\
 \delta v(\vec{r}) &=& \cos(\vec{b}_1\! \cdot \! \vec{r})
+ \cos(\vec{b}_2 \! \cdot \! \vec{r}),
\end{eqnarray}
where $v(\vec{r})$ is given by (\ref{eq:honeypot}).
Note that when only one field strength is imbalanced, the corresponding
potential still displays a reflection symmetry.
In the present case, it is the $Ox$-reflection symmetry because $V'(\vec{r})$
is invariant under the exchange $\vec{b}_1 \leftrightarrow \vec{b}_2$.
Requiring now that the primitive cell $\Sigma$ exhibits two field minima
imposes $-1 \leq \eta \leq 1$.
Their positions in $\Sigma$ are given by $\vec{r'}_{\textsc{a},\textsc{b}} = \varphi_{\textsc{a},\textsc{b}} \,
(\vec{a}_1 + \vec{a}_2)$ with $\cos(2\pi\varphi_{\textsc{a},\textsc{b}}) = -(1+\eta)/2$.
Their mid-point $\vec{r'}_\textsc{s} = (\vec{r'}_\textsc{a}+\vec{r'}_\textsc{b})/2 = (\vec{a}_1 +\vec{a}_2)/2$ is a saddle point and defines the potential barrier height
$V'_\textsc{s}$ to cross to go from $\textsc{a}$ and $\textsc{b}$ in $\Sigma$.
One finds $V'_\textsc{s}= (\eta-1)^2 V_0$.

\begin{figure}[tb]
\includegraphics[scale=0.7]{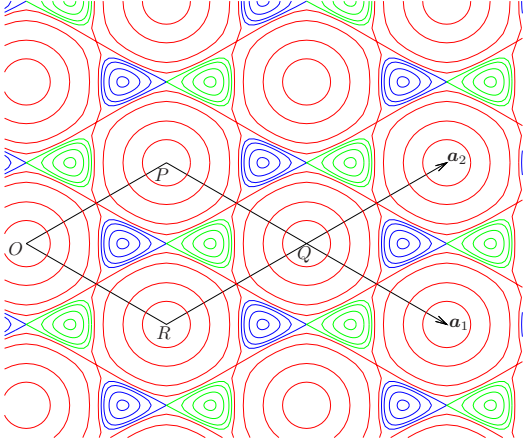}
\caption{\label{AsymLatticeRc0.81}
[Color online] Slightly distorted lattice obtained with vanishing mismatch
angles and one imbalanced field strength, namely $s_1 \!=\! 10/9$ and $s_2
\!=\! s_3 \!=\! 1$.
In this particular case the hexagon of field minima is slightly squeezed along
the horizontal axis $Ox$ and the vectors $\vec{c'}_n$ connecting a given
minimum to its three nearest-neghbors have now different lengths.
In the situation depicted $\biglb|\vec{c'}_2\bigrb| \!=\! \biglb|\vec{c'}_3\bigrb| \!\not=\!
\biglb|\vec{c'}_1\bigrb|$.
In turn, due to the reflection symmetry about $Ox$, the tight-binding hopping
amplitudes satisfy $\biglb|t_2\bigrb| \!=\! \biglb|t_3\bigrb| \!\not=\! \biglb|t_1\bigrb|$. 
}
\end{figure}

As a whole the field minima organize in a hexagon which is stretched ($\eta$
negative) or compressed ($\eta$ positive) along $Ox$, see
Fig.~\ref{AsymLatticeRc0.81}.
As a consequence two of the three new vectors $\vec{c'}_n$ joining one minimum
to its three nearest-neighbors will have equal length.
In the present situation we get $\biglb|\vec{c'}_2\bigrb| = \biglb|\vec{c'}_3\bigrb| \not=
\biglb|\vec{c'}_1\bigrb|$.
The potential barrier height $V''_\textsc{s}$ to cross to go from $\textsc{a}$ to $\textsc{b}$ along
$\vec{c'}_2$ and $\vec{c'}_3$ is given by the corresponding saddle points
located at the middle of the edges of $\Sigma$.
One finds $V''_\textsc{s} = (\eta+1)^2V_0$.

Now, when $\eta$ is increased from $0$, the minima move closer along
$\vec{c'}_1$ and move away along $\vec{c'}_2$ and $\vec{c'}_3$.
At the same time, the potential barrier $V'_\textsc{s}$ along $\vec{c'}_1$ is lowered
and the the potential barrier $V''_\textsc{s}$ along $\vec{c'}_2$ and $\vec{c'}_3$ is
increased.
As a net effect, in the tight-binding picture, we expect the tunneling
amplitude $\biglb|t_1\bigrb|$ to increase while $\biglb|t_2\bigrb|$ and $\biglb|t_3\bigrb|$ decrease.
We get the opposite conclusion when $\eta$ is lowered from $0$. Since the
potential is invariant through $\vec{b}_1 \leftrightarrow \vec{b}_2$, we
further have $\biglb|t_2\bigrb|=\biglb|t_3\bigrb|$ and we recover the case of one imbalanced hopping
amplitude analyzed in the previous section.

One could try to derive a semiclassical expression of the $t_n$ as a function
of $\eta$ using the instanton method but, actually, such a tedious calculation
proves unnecessary, at least when $\eta$ is small.
Indeed, by inspection of the semiclassical expression \eqref{semi-t}, we
expect the ratio $\biglb|t_1/t_2\bigrb|$ to scale as $\exp(\Delta S(\eta)/\hbar_e)$ at
leading order, where $\Delta S(\eta)$ is the action difference between the two
instanton trajectories linking sites $\textsc{a}$ and $\textsc{b}$ along $\vec{c'}_2$ and
$\vec{c'}_1$ respectively.
For small enough $\eta$ we expect $\Delta S(\eta)$ to grow linearly with
$\eta$, the slope being positive since the ratio $\biglb|t_1/t_2\bigrb|$ should increase
with $\eta$.
The Dirac degeneracies disappear when this ratio is $2$ (see previous
section), thus we get the semiclassical prediction that this will happen when
$\eta\propto\hbar_e$.
This result can also be inferred by saying that the Dirac points will
disappear as soon as the perturbing potential $2 \eta \delta V(\vec{r})$, see
\eqref{new-v}, strongly mixes the unperturbed states.
This will happen when the corresponding coupling energy equals the mean level
spacing of the unperturbed system and we get back to the prediction $\eta\propto\hbar_e$.

\begin{figure}[tb]
\includegraphics[scale=0.7]{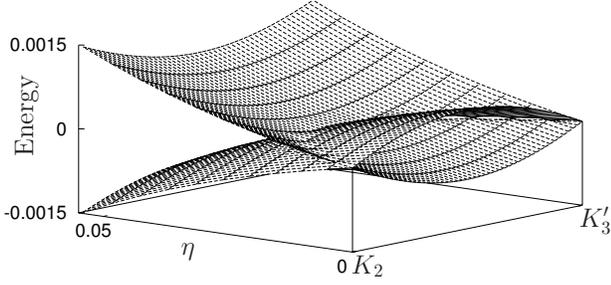}\caption{\label{fig:bandr}
The band diagram for the two lowest levels as a function of $\eta$ for
$V_0=80E_R$ ($\hbar_e \approx 0.158)$.
The bands are plotted along the vertical straight line joining the Dirac
points $K_2$ and $K'_3$ of the balanced situation, see
Fig.~\ref{fig:honeycombRecipLattice}.
The origin of energy is fixed at the Fermi energy for a half-filled band and
all bands have been shifted such that the upper and lower bands intersect at
zero energy difference. }
\end{figure}

\begin{figure}[tb]
\includegraphics[scale=0.9]{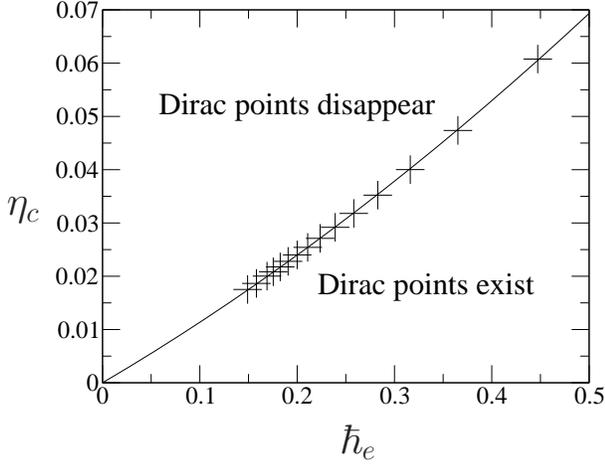}
\caption{\label{fig:rc} The critical laser strength imbalance $\eta_c$ at
which the Dirac degeneracies are lifted as a function of the effective
Planck's constant $\hbar_e = \sqrt{2E_R/V_0}$.
The solid line corresponds to a quadratic fit of the numerical data.
The linear coefficient is $\alpha \approx 0.1074$ while the quadratic one is
$\beta \approx 0.0624$.
As one can see our numerical results are in good agreement with our
semiclassical prediction $\eta_c \propto \hbar_e$.
The degree of control of the intensity imbalance of the laser fields gets more
stringent as the optical lattice depth $V_0$ is increased.
Nevertheless, at already $V_0 = 20 E_R$ ($\hbar_e \approx 0.3$), the laser
intensities should all equal within $8\%$ which does not sound particularly
demanding.}
\end{figure}

To check our semiclassical prediction we have computed, for each value of the
effective Planck's constant $\hbar_e$, the ground state and first excited-state
levels for different values of $\eta$ and we have extracted the corresponding
critical value $\eta_c$ for which the Dirac degeneracies are lifted.
Figure~\ref{fig:bandr} gives an example of the band structure obtained at
$\hbar_e = 1/\sqrt{40} \approx 0.16$ for $\eta$ ranging from $0$ to $0.054$.
We have then plotted $\eta_c$ as a function of $\hbar_e$, see Fig.~\ref{fig:rc}.
We have fitted the data with the quadratic fit function {$\alpha \hbar_e +
\beta \hbar^2_e$} and found $\alpha\approx0.1074$ and $\beta\approx0.0624$
enforcing the very good agreement obtained with our linear prediction in the
semiclassical regime $\hbar_e \ll 1$.
The quadratic correction could certainly be inferred from semiclassical
higher-order corrections.

We would like to emphasize at this point that increasing or decreasing $\eta$
from $0$ is not symmetrical.
When $\eta$ is decreased from $0$, the Dirac degeneracies are predicted to
disappear when $\biglb|t_1/t_2\bigrb| \to 0$.
However the best that we can do is to let $\eta \to -1$.
This unfortunately means that one laser beam is almost extinguished and the
situation is more that of very weakly coupled one-dimensional chains, a
situation we postpone to future study as it proves interesting for high-$T_c$
superconductivity~\cite{lee2006}.
We thus see that decreasing slightly $\eta$ from $0$ does not harm the Dirac
degeneracies.
They move inside $\Omega$ but do survive.
By contrast, increasing slightly $\eta$ from $0$ does destroy the Dirac
degeneracies as soon as $\eta \sim \hbar_e$.

As one can see from the plots, the tolerance about the intensity mismatch of
the laser beams increases with $\hbar_e$, or equivalently when the optical
lattice depth $V_0$ decreases.
On the other hand, as we already saw, the Dirac cones do not extend much over
the Brillouin zone if $V_0$ is too small.
So there is a trade-off to make. The situation is however really favorable
since the intensity mismatch tolerance is already in the $10\%$ range for $V_0
\sim 10 E_R$.
This means that the massless Dirac fermions prove quite robust and should be
easily accessed experimentally.

\subsubsection{Critical in-plane angle mismatch}

\begin{figure}[tb]
\includegraphics[scale=0.65]{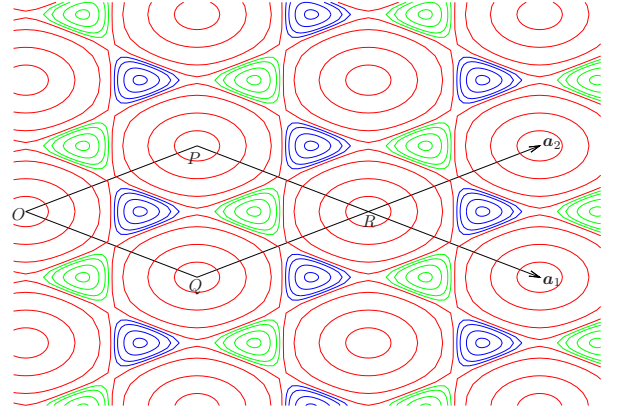}
\caption{\label{Thetac0.1pi.eps}
[Color online] Distorted lattice obtained with balanced field strengths $s_n
\!=\! 1$ and angle mismatch $\theta_3\!=\! -\theta_2 \!=\! -\pi/10$.
In this particular case the hexagon of field minima is stretched along the
horizontal axis $Ox$ and the vectors $\vec{c'}_n$ connecting a given minimum
to its three nearest-neghbors have now different lengths.
In the situation depicted $\biglb|\vec{c'}_2\bigrb| \!=\! \biglb|\vec{c'}_3\bigrb| \!\not=\!
\biglb|\vec{c'}_1\bigrb|$.
In turn, due to the reflection symmetry about $Ox$, the tight-binding hopping
amplitudes satisfy $\biglb|t_2\bigrb| \!=\! \biglb|t_3\bigrb| \!\not=\! \biglb|t_1\bigrb|$. 
}
\end{figure}

We now estimate the critical angle mismatch when all laser beams have
the same intensities ($s_1=s_2=s_3=1$).
We see from \eqref{newfield} that the new optical potential still
displays the exchange symmetry $\vec{b'}_1 \leftrightarrow \vec{b'}_2$ and
thus a reflection invariance with respect to their bisectrix.
In the following we stick to the simple case where $\theta_3 = -\theta_2
=\theta$ and $\theta$ is small.
In this case both the Bravais lattice, the reciprocal lattice, the Brillouin
zone $\Omega$ and the diamond-shaped primitive cell $\Sigma$ get modified.
The new reciprocal basis vectors turn out to be $\vec{b'}_1 = \vec{b}_1 +
\boldsymbol{\delta}\vec{b}_1$, $\vec{b'}_2 = \vec{b}_2 +
\boldsymbol{\delta}\vec{b}_2$ where $\boldsymbol{\delta}\vec{b}_1 =
(\theta/\sqrt{3}) \, \vec{b}_2$ and $\boldsymbol{\delta}\vec{b}_2 =
(\theta/\sqrt{3}) \, \vec{b}_1$.
Since the exchange symmetry $\vec{b}_1 \leftrightarrow \vec{b}_2$ is again
preserved, the new potential continues to display the $Ox$-reflection invariance.
Figure~\ref{Thetac0.1pi.eps} gives a plot of the new potential structure for
$\theta = -\pi/10$.

This situation boils down again to the case of one imbalanced tunneling
amplitude.
Indeed, the angle between the $\vec{b'}_1$ and $\vec{b'}_2$ decreases when $\theta$ is
increased from $0$.
In turn the angle between the corresponding $\vec{a'}_n$ increases and the
hexagon structure made by the $\textsc{a}$ and $\textsc{b}$ minima get compressed along $Ox$.
The opposite conclusion holds when $\theta$ is decreased from $0$.
We get again the situation where $\biglb|t_2\bigrb| = \biglb|t_3\bigrb| \not= \biglb|t_1\bigrb|$ and $\biglb|t_1/t_2\bigrb|
\geq 1$ when $\theta \geq 0$ and vice-versa.
Like for the field strength imbalance, the situations $\theta >0$ and $\theta
<0$ are not symmetric.
The masless Dirac fermions prove more sensitive to {\it closing} the angle
between the $\vec{b'}_n$, so for $\theta_3 = -\theta_2 = \theta >0$ because
$\biglb|t_1/t_2\bigrb|$ then increases and the threshold $\biglb|t_1/t_2\bigrb|=2$ is more rapidly
hit.
This is the situation we explore.

\begin{figure}[tb]
\includegraphics[scale=0.9]{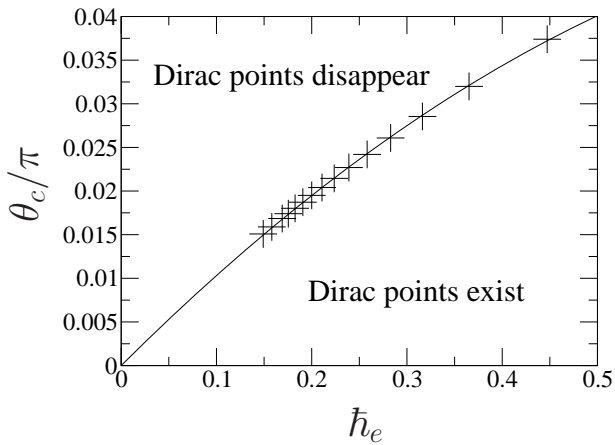}
\caption{\label{fig:thetac}
The critical angle mismatch $\theta_c$ (in units of $\pi$) beyond which the
Dirac degeneracies disappear as a function of the effective Planck's constant
$\hbar_e = \sqrt{2E_R/V_0}$.
The dashed line corresponds to a quadratic fit of the numerical data.
The linear coefficient is $0.109$ while the quadratic one is $-0.0577$.
As one can see our numerical results are in good agreement with our
semiclassical prediction $\theta_c \propto \hbar_e$.
The degree of control of the angle mismatch gets more stringent as the optical
lattice depth $V_0$ is increased.
Nevertheless, at already $V_0 = 20 E_R$ ($\hbar_e \approx 0.3$), the angle
mismatch should be less than $5^\circ $ which is not particularly demanding.}
\end{figure}

Applying the same reasoning as before, we thus predict the critical angle
mismatch beyond which the massless Dirac fermions are destroyed to scale as
$\theta_c \propto \hbar_e$.
Again, to get $\theta_c$ as a function of $\hbar_e$, we numerically compute
the band structure at a given $\hbar_e$ for different in-plane mismatch angles
$\theta$ and then extract the value $\theta_c$ for which the Dirac degeneracy
is lifted.
We then repeat the procedure for different $\hbar_e$.
As one can see, our prediction is in very good agreement with the numerical
calculations, see Fig.~\ref{fig:thetac}, and well supported by a quadratic
fit.
As $\theta_c$ increases with $\hbar_e$, there is a trade-off to make between reaching the tight-binding regime where $V_0$ is large and achieving an experimentally reasonable angle mismatch tolerance which requires $V_0$ to be small.
The trade-off turns out to be a favorable one since already for $V_0=20E_R$
($\hbar_e \approx 0.3$), one gets a tolerance of about $5^\circ $ on the laser
beams alignment.
We expect the same type of scaling for small out-of-plane angle mismatches.
Furthermore, when several small distortions combine, their effects should add
up and thus the critical imperfection threshold should still scale with
$\hbar_e$.

As an overall conclusion we see that massless Dirac fermions are quite robust
to moderate lattice distortions.
Demonstrating them in an experiment should not be particularly demanding in terms of the control of the laser configuration.

\subsection{Inequivalent potential wells}\label{sec:V_phi}

We finally briefly mention how to distort the optical lattice in a systematic manner as it allows for an experimental control of the mass of the Dirac fermions as well as for a continuous switch from a honeycomb lattice to a triangular one.

In Sec.~\ref{sec:HCvsGRAPH}, we observed that the honeycomb potential
\eqref{eq:HCpotential} is the simplest of all graphene-type potentials,
characterized by choosing $v_{\vec{0}}$ and ${v_{\vec{b}_1}}$ real
(in fact, positive) while putting all unrelated coefficients in
\eqref{eq:graphtype} to zero.
Now, letting $v_{\vec{b}_1}$ to acquire a phase $\varphi$, such that
$\Exp{-\mri\varphi}v_{\vec{b}_1}$ is positive, will break the reflection symmetry of the honeycomb potential \cite{followup}.

In the $\vec{r}$-dependent part of the dimensionless potential
\eqref{eq:honeypot}, this phase $\varphi$ is introduced by the replacement
\begin{equation}
  \label{eq:v-phi}
  \sum_{a=1}^3\cos(\vec{b}_a\cdot\vec{r})\to\sum_{a=1}^3\cos(\vec{b}_a\cdot\vec{r}+\varphi)
\,,
\end{equation}
where $\vec{b}_3=-\vec{b}_1-\vec{b}_2$. This can be implemented by superimposing three \emph{independent} standing
waves, of the same wavelength and with equal intensity, whose wave vectors
form the trine of Fig.~\ref{fig:threeBeam}~\cite{David}.
As a consequence of the incoherent superposition, the $t$ replacement of
\eqref{eq:jointshift} is not available, and the $\vec{r}$ replacement alone
cannot remove all three phases of the standing waves.
One can, however, shift $\vec{r}$ such that the three phases are the same, and
then one has an intensity pattern proportional to the right-hand side of
\eqref{eq:v-phi}.

Most of the hexagon structure of Fig.~\ref{fig:ABChexagon} remains unchanged
by this modification: lattice sites $\textsc{a}$, $\textsc{b}$, $\textsc{c}$
continue to be the locations of local minima and maxima, whereas the saddle
points \textsc{s} acquire new positions on the \dots\textsc{abcabc}\dots\
lines.

\begin{figure}[tb]
\centerline{\includegraphics{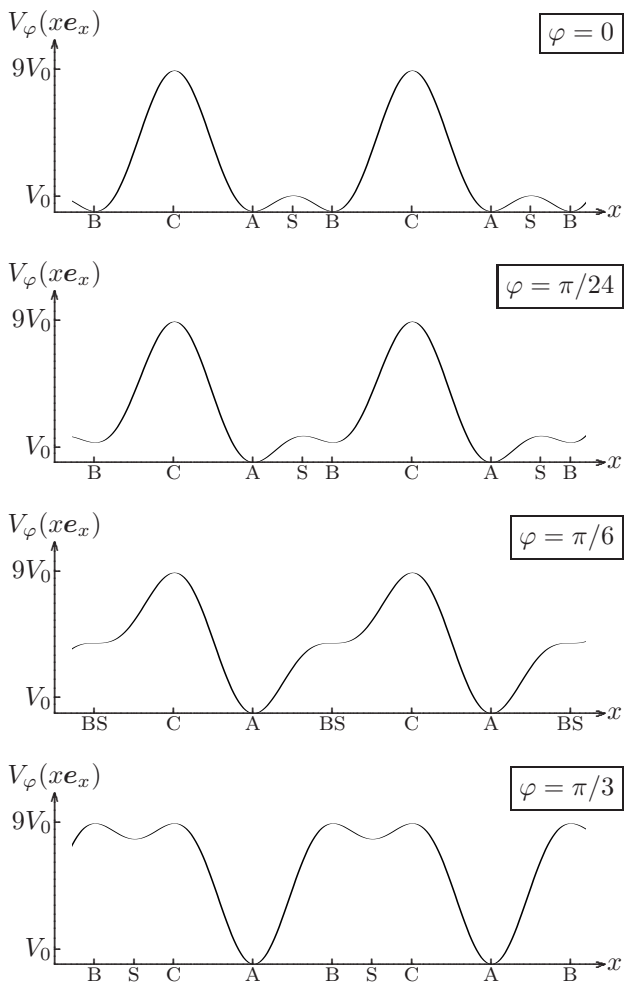}}
\caption{\label{fig:ABCSphi}%
For various values of the phase parameter $\varphi$ of \eqref{eq:v-phi}, the
plot shows the potential energy along a \dots\textsc{abcabc}\dots\ line in
Fig.~\ref{fig:ABChexagon}.
The top plot, for ${\varphi=0}$, repeats the bottom-left plot of
Fig.~\ref{fig:ABChexagon} for reference.
The degeneracy between sites \textsc{a} and \textsc{b} is lifted for the small
$\varphi$ value of ${\varphi=\pi/24}$, the saddle points have moved closer to
the \textsc{b} sites, where we continue to have local minima. In this situation the Dirac fermions acquire a mass ${m_*\propto\biglb|\varphi\bigrb|}$.
When ${\varphi=\pi/6}$, the saddle points \textsc{s} coincide with the
\textsc{b} sites, and we have cubic saddle points there.
Finally, in the bottom plot, we have ${\varphi=\pi/3}$ and the saddle points
are halfway between adjacent \textsc{b} and \textsc{c} sites, with potential
maxima at both of them.
Except for a displacement, the potential in the bottom plot is the negative of
the potential in the top plot, and thus identical with the honeycomb potential
\eqref{eq:HCpotential} for red rather than blue detuning of the three running
wave lasers.
For ease of comparison, the potential constants are adjusted such that the
maxima and minima are at ${V=0}$ and ${V=9V_0}$, respectively, for all
$\varphi$ values.
}
\end{figure}

Figure~\ref{fig:ABCSphi} confirms that, for small $\varphi$ values, the minima
of the honeycomb dipole potential are still organized in a hexagonal pattern
but we now have \emph{different} potential depths at sites \textsc{a} and
\textsc{b}. The potential energy mismatch is $2\epsilon \approx 8V_0\biglb|\varphi\bigrb|/\sqrt{3}$. In view of \eqref{eq:bs1} and \eqref{eq:bs2}, this means that the Dirac fermions acquire a mass ${m_*\propto\biglb|\varphi\bigrb|}$ or, in other words, that the Dirac
degeneracies are lifted. The possibility of fine-tuning the mass of the Dirac fermions through the parameter $\varphi$ is an interesting experimental knob to play with.

Increasing ${\biglb|\varphi\bigrb|}$ further, one can also see that, for the particular values ${\biglb|\varphi\bigrb|}=\pi/6$ and $\pi/2$, the three sublattices of saddle points merge into a single
triangular lattice, which coincides with the \textsc{a}, \textsc{b}, or
\textsc{c} lattice, respectively; see Fig.~\ref{fig:ABCSphi}.
This merging of a potential minimum or maximum with three saddle points, leads
to a peculiar third-order saddle point. For $\varphi=\pi/6$, say, the \textsc{s} sites merge with the \textsc{b} sites
and we have
\begin{equation}
  \label{eq:mergeSandB}
  \sum_a\cos(\vec{b}_a\cdot\vec{r}+\varphi)
\Bigr|_{\varphi=\pi/6}
\approx
-\frac{1}{6}\sum_a\bigl[\vec{b}_a\cdot(\vec{r}-\vec{r}_{\textsc{b}})\bigr]^3
\end{equation}
for ${\biglb|\vec{r}-\vec{r}_{\textsc{b}}\bigrb|\ll\kappa^{-1}}$,
hence a cubic saddle point rather than the usual quadratic saddle point.

An unpolarized ultracold gas of spin-$\frac{1}{2}$ fermions loaded into
such a potential at half-filling would lead to two fermions per well.
By driving the system through attractive interactions, one could even get a
Mott insulator of fermion pairs.
By switching off all interactions and setting ${\varphi=0}$,
one should be able to study oscillations of atoms between the \textsc{a} and
\textsc{b} sublattices.
We will analyze this situation in a follow-up paper.

\section{Conclusion}

Motivated by the vivid field of graphene physics, we have explained and
analyzed how to reproduce massless Dirac fermions by loading ultracold
fermions in an optical lattice with honeycomb structure.
We have described the two-dimensional laser configuration that gives rise to an
optical potential where field minima are organized in a honeycomb structure
(with lattice constant $a$) and we have thoroughly detailed the corresponding
crystallographic features.
The behavior of atoms propagating in such an optical potential in the
tight-binding regime is in one-to-one correspondence with the behavior of
electrons propagating in a graphene sheet.
The ground state and first-excited levels of the band structure exhibit two
conical degeneracies located at the corners of the first Brillouin zone, as
dictated by symmetry arguments.
In the neighborhood of these degeneracies, the band spectrum is linear.

When
the lattice is loaded with fermions at half-filling, the Fermi energy slices
the band structure at these degeneracy points, known as the Dirac
points. Around half-filling, the tight-binding Hamilton operator can then be recast
in a form reminiscent of the relativistic Weyl-Dirac Hamilton operator and featuring
so-called massless Dirac fermions.
The important parameter driving the dynamics turns out to be the hopping
amplitude $t_0$ between nearest-neighbors sites as it gives the band width
$W=6\biglb|t_0\bigrb|$ and the ``Fermi velocity'' $v_0= 3a\biglb|t_0\bigrb|/(2\hbar)$.
We have derived a semiclassical expression for $\biglb|t_0\bigrb|$ in terms of the
effective Planck's constant of the problem, namely $\hbar_e = \sqrt{2E_R/V_0}$
(with $V_0$ the optical potential strength and $E_R$ the recoil energy) and
have compared it to an exact numerical calculation of the band spectrum.
From this we have derived quantitative experimental criteria (such as the
required initial temperature of the atomic gas) to reach the massless Dirac
fermions regime.

We have also examined the robustness of the massless Dirac fermions to
imperfections of the laser configuration (field strengths imbalances and angle
mismatches).
Massless Dirac fermions turn out to be quite robust as the equality of the
beam intensities should be controlled  within the few percent range while the
respective beam angles should equal $2\pi/3$ within the few degrees range.
By appropriately controlling these lattice distortions, one can even control
and move the Dirac points in the Brillouin zone. Lastly, we introduce an irremovable phase to the honeycomb potential, hence lifting the degeneracy between two sublattices. In turn, the Dirac fermions acquire a mass proportional to this phase. We also briefly mention the peculiar properties of saddle points and the possibility to study oscillations of atoms between two sublattices as a consequence of this irremovable phase.

As an overall conclusion, mimicking graphene physics with ultracold fermions
is within experimental reach.
For non-interacting fermions, one could think of implementing transport
experiments (in the presence of disorder or not).
For example, by rotating the whole honeycomb lattice around a given axis
perpendicular to the lattice plane \cite{tung06} or by implementing the scheme
proposed in \cite{soerensen05}, one would mimic effective magnetic fields able
to reproduce the quantum Hall effect situation. In the rest frame of the
atoms, the centrifugal effects are described by a fictitious vector
potential.
This leads to Landau levels and paves the way to physical effects analogous to
the quantum Hall effect. The possibilty to move the Dirac points in the
Brillouin zone even offers new physical effects to test \cite{dietl08}.

Interacting systems on a lattice prove also particularly interesting as they
can be mapped (at least for strong interactions) on Heisenberg models and thus
offer ways of exploring quantum magnetism~\cite{demler2004}.
In the case of the honeycomb lattice, quantum phase transitions are predicted
to occur when the interaction strength $\biglb|U\bigrb|$ is strong enough.
For repulsive interactions, quantum Monte-Carlo calculations predict anti-ferromagnetic order to occur at half-filling
\cite{paiva05}.
For attractive interactions, mean-field calculations have started to analyze
the BEC-BCS crossover and predict a semi-metal/superconductor transition
\cite{zhao06}.
Recent Monte-carlo studies have even started to analyze this BEC-BCS crossover
\cite{su09} and one can expect an increase of such studies in the near
future.
Very recently, implementations of massless Dirac fermions in square lattices
have been proposed \cite{hou09, goldman09}.
The situation seems thus mature for an experimental effort towards loading
ultracold fermions in a honeycomb optical lattice.

\begin{acknowledgments}
BG and ChM would like to thank Dominique Delande, Gilles Montambaux, Jean-No\"el Fuchs, Mark
Goerbig, and
David Wilkowski for stimulating discussions.
LKL acknowledges support from the French Merlion-PhD programme (CNOUS
20074539).
This work has also been supported by the CNRS PICS No.~4159
(France) and the France-Singapore Merlion programme (SpinCold 2.02.07).
Centre for Quantum Technologies is a Research Centre of Excellence funded by
Ministry of Education and National Research Foundation of Singapore.
\end{acknowledgments}

\bibliography{UcFGtLv05}

\begin{thebibliography}{66}
\bibitem[{fol()}]{followup}
\bibinfo{note}{The symmetry aspects of honeycomb potential will be discussed in
  more details in a follow-up paper.}
\end{thebibliography}


\end{document}